\documentclass[pre,amsmath,amssymb,superscriptaddress,showpacs]{revtex4-1}

\usepackage{amsmath,amssymb,graphicx}
\usepackage{amsbsy}
\usepackage{latexsym}
\usepackage{color}
\usepackage{graphicx}
\usepackage{psfrag}
\usepackage[normalem]{ulem}
\usepackage{bm}

\newcommand{\be}{\begin{equation}}
\newcommand{\ee}{\end{equation}}
\newcommand{\bea}{\begin{eqnarray}}
\newcommand{\eea}{\end{eqnarray}}

\newcommand{\comment}[1]{}

\begin{document}

\title{Kinetics of the Two-dimensional Long-range Ising Model at Low Temperatures}

\author{Ramgopal Agrawal}
\email{ramgopal.sps@gmail.com}
\affiliation{School of Physical Sciences, Jawaharlal Nehru University, New Delhi 110067, India.}
\author{Federico Corberi}
\email{corberi@sa.infn.it}
\affiliation{Dipartimento di Fisica ``E.~R. Caianiello'', and INFN, Gruppo Collegato di Salerno,
and CNISM, Unit\`a di Salerno, Universit\`a  di Salerno, via Giovanni Paolo II 132, 84084 Fisciano (SA), Italy.}
\author{Eugenio Lippiello}
\email{eugenio.lippiello@unicampania.it}
\affiliation{Dipartimento di Matematica e Fisica, Universit\`a della Campania,
Viale Lincoln 5, 81100, Caserta, Italy}
\author{Paolo Politi}
\email{paolo.politi@cnr.it}
\affiliation{Istituto dei Sistemi Complessi, Consiglio Nazionale delle Ricerche, Via Madonna del Piano 10, 50019 Sesto Fiorentino, Italy}
\affiliation{INFN Sezione di Firenze, via G. Sansone 1, 50019 Sesto Fiorentino, Italy}
\author{Sanjay Puri}
\email{purijnu@gmail.com}
\affiliation{School of Physical Sciences, Jawaharlal Nehru University, New Delhi 110067, India.}

\begin{abstract}
  We study the low-temperature domain growth kinetics of the two-dimensional Ising model with long-range coupling: $J(r) \sim r^{-(d+\sigma)}$, where $d=2$ is the dimensionality.
  According to the Bray-Rutenberg predictions, the exponent $\sigma$ controls the algebraic  growth in time of the characteristic domain size $L(t)$, $L(t) \sim t^{1/z}$, with growth exponent $z=1+\sigma$ for $\sigma <1$ and $z=2$ for $\sigma >1$. These results hold for quenches to a non-zero temperature $T>0$ below the critical temperature $T_c$.
  We show that, in the case of quenches to $T=0$, due to the long-range interactions, the interfaces experience a drift which makes the dynamics of the system peculiar. More precisely we find that in this case the  growth exponent takes the value $z=4/3$, independent of $\sigma$, showing that it is a universal quantity. We support our claim by means of extended Monte Carlo  simulations and analytical arguments for single domains.
  
\end{abstract}

\maketitle

\section{Introduction}
\label{intro}

After a quench from the disordered phase above the critical temperature
$T_c$ to a final temperature $T<T_c$ ferromagnetic materials undergo 
phase-ordering~\cite{PuriWad09,Bray94,CRP,Cor15,CorCugYos11}.
The system orders locally inside domains whose typical size $L(t)$ grows in time 
until equilibration takes place on a timescale $\tau _{eq}$ which diverges with the system (linear) size ${\cal L}$.
This relaxation process is called coarsening and it is often accompanied by a dynamical scaling symmetry,
which amounts to the physical fact that configurations at different times are statistically similar
upon measuring distances in unit of $L(t)$. The latter usually increases algebraically,
$L(t)\sim t^{1/z}$, where $z$ is a non-equilibrium dynamical exponent that is unrelated to
any equilibrium property. This exponent is also independent of the quench temperature~\cite{Bray94,Bray90_rg}, a property
which is true for any universal quantity, because it can
be shown that temperature is an irrelevant parameter in the sense of the renormalisation group~\cite{Bray90_rg}.
This statement applies for quenches to $0<T<T_c$: indeed when cooling down to $T=T_c$
the process is qualitatively different because the order parameter vanishes in the target equilibrium state, at variance with 
what happens when $T<T_c$. Quenches to $T=0$, on the other hand,
may also have peculiar properties because any activated process is forbidden. 
To be concrete, let us discuss a system with a scalar non-conserved order parameter. 

{\bf Short-range systems --} If interactions are restricted to nearest
neighboring (nn) spins, the Ising model with Glauber single spin-flip kinetics~\cite{Glauber1963} 
represents an appropriate description.  
Letting space dimension $d>1$ above the lower critical one, in order to have a finite $T_c$,
after quenching to $0<T<T_c$ ordered domains grow at late times with $z=z_{\rm cd}=2$ until the system eventually
attains the equilibrium state in a time $\tau_{eq}$ which
is in most cases $\tau _{eq} \sim {\cal L}^z$. We indicate with $z_{\rm cd}$ this value of $z$ because the
motion of interface in this case is curvature driven~\cite{AllenCahn79,Bray94}.

\begin{figure}[h]
\includegraphics[width=0.8\textwidth]{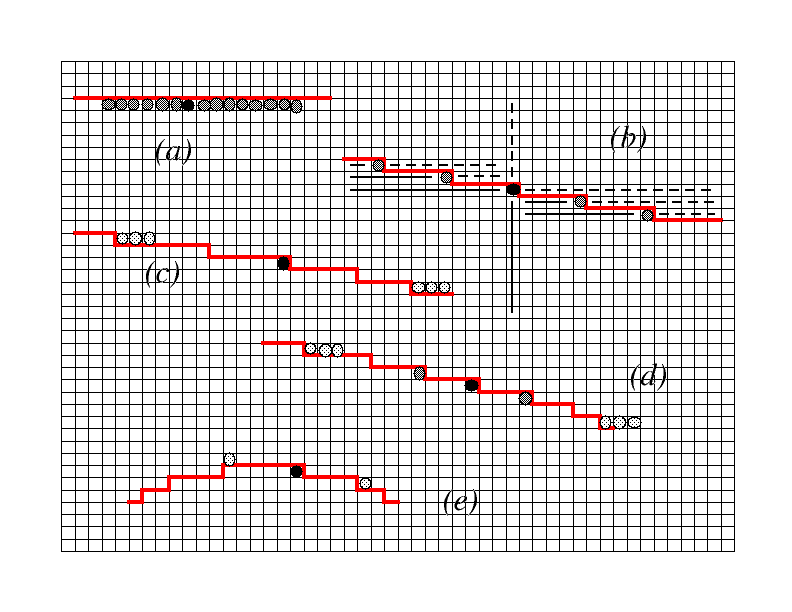}
\caption{Pictorial representation of different interfaces (thick red lines) on a square lattice
and their stability with long-range interactions. 
Spins are located in the centre of the small squares forming the lattice and
they have opposite orientation on the two sides of the interface, supposed to extend indefinitely.
We show: (a) a flat interface directed along one lattice direction;
(b) an interface with a constant slope (i.e. all steps have the same length);
(c) an interface of positive curvature (i.e. steps become shorter upon going right);
(d) an interface with a local constant slope, then acquiring a finite curvature 
(the three central terraces have the same length, higher terraces have an increasing length,
lower terraces have a decreasing length); 
(e) an interface whose slope changes sign.
For any shape we focus on a given spin on the interface (full black circle $\bullet$) and we wonder about its
stability at $T=0$. Using symmetry considerations it is possible to compensate the interaction
with spins of opposite orientations. This is explicitly depicted in (b) where the interaction
with spins parallel to $\bullet$ (straight segments) is compensated with
the interaction with spins anti-parallel to $\bullet$ (dashed segments). 
In particular, all spins within the two perpendicular straight segments originating in $\bullet$ are
compensated by all spins within the two perpendicular dashed segments originating in the same site.
Once this procedure is completed, uncompensated spins are shown as dark grey circles
if they are parallel to $\bullet$ (therefore, they are stabilising)
and as light grey circles 
if they are anti-parallel to $\bullet$ (therefore, they are destabilising).
Therefore, spins $\bullet$ at interfaces (a) and (b) are stable while spin $\bullet$ in (c) is unstable.
Notice that all the above is true for any value of $\sigma$.
The stability of spin $\bullet$ in (d) is more complicated because it has short-distance
stabilising interactions (dark grey circles) and longer-distance destabilising interactions
(light grey circles). The resulting effect depends on the details of the interface and on $\sigma$:
for large $\sigma$ we expect the short-distance stabilising interactions to prevail while
for small $\sigma$ we expect the longer-distance destabilising interactions to do so.
Finally, in (e) the edge spins of a top terrace are unstable showing that curvature is more
relevant than the sign of the slope.
In all figures (b-e) the spin at the right of $\bullet$ (i.e., the spin on the other side of the interface)
is stable. 
For the nn model the picture is different because interaction is local and the curvature is irrelevant:
interface (a) is stable while edge spins in (b-e) are (energetically) in neutral equilibrium,
therefore are dynamically unstable, see Eq.~(\ref{metrop}) with $\Delta E =0$.}
\label{stab_interf}
\end{figure}

For quenches to $T=0$, even if the motion of interfaces is no lomger curvature driven but has a diffusive character,
one still observes a growth law with an exponent $z=z_{\rm diff}=z_{\rm cd}=2$. 
Another difference between $T=0$ and $T>0$ is that the equilibration in the former quench 
may be impeded due to blocking of the system into infinitely lived 
metastable states~\cite{BarKraRed09,BlaCorCugPic14}. In $d=2$ these blocked states are stripes 
with flat interfaces extending along the lattice directions, as the one denoted with the letter $a$ in Fig.\ref{stab_interf}. 
Clearly, the fact that such
flat interfaces are frozen does not only affect the fate of the system in a quench
to $T=0$, but also the preceding dynamics. Indeed it was shown in~\cite{CorLipZan08}
that, although the value of $z$ does not change in going from $T>0$ to $T=0$, since $z_{\rm diff}=z_{\rm cd}$,
some other non-equilibrium exponents related to the geometry of interfaces, do change. 

Furthermore,
even if the quench is made to a finite $T$, the different dynamics associated to $T=0$ is observed
in a pre-asymptotic regime which can be rather long if $T$ is small enough. 
Similarly, for sufficiently low $T>0$, although the metastable states are eventually escaped, this happens on 
timescales that can be huge, greatly delaying the equilibration with respect to what happens at 
higher $T$ where, as already mentioned, $\tau_{eq}\sim {\cal L}^z$.  
Summarising, the case with $T=0$ displays peculiar features which can strongly affect even the dynamics
of quenches to finite $T$, at least pre-asymptotically. 

A separated discussion is deserved by the case $d=1$, because here $T_c=0$, meaning that, strictly speaking $T=0$ is the only possibility to cool the system to a magnetised state. However, even for quenches to a finite $T$, a coarsening stage takes 
place because domains keep growing as in a $T=0$ quench until their size $L(t)$ reaches the equilibrium
correlation length $\xi$ or the system size ${\cal L}$, after which equilibration occurs. Also in this case the motion of
interfaces has a diffusive character and one finds $z=z_{\rm diff}=2$~\cite{Glauber1963}
but, at variance with higher dimensions, there are no metastable states due to the trivial $1d$ lattice
geometry. 
  
{\bf Long-range systems --} The presence of long-range interactions changes a lot the above picture.
The equilibrium scenario
with a coupling between spins at distance $r$ decaying as $r^{-(d+\sigma)}$ 
is well understood. There is an ordered phase
below a finite $T_c$ also in $d=1$ provided, in this case, that $\sigma \le 1$.  
Furthermore, the energy is extensive and the system is additive for any $d$ if $\sigma >0$, a regime sometimes
denoted as weak long-range, whereas extensivity and additivity are lost for $\sigma \le 0$, the strong long-range case. 
In this paper we will only focus on the weak long-range case $\sigma >0$.  

Regarding the non-equilibrium properties, in quenches to a finite $T$ one finds~\cite{BrayRut94,CMJ19,PhysRevLett.125.180601}
$z=z_{\rm adv}=1+\sigma$ for $\sigma \le1$ and  $z=2$ for $\sigma >1$ (with logarithmic corrections right at $\sigma =1$).
This applies down to $d=1$ where, if $\sigma >1$, one has $T_c=0$ and the discussion made above for short-range
interactions regarding equilibration applies. 
The different behaviour of the exponent $z$ in crossing $\sigma=1$ can be ascribed to qualitatively different underlying dynamical mechanisms.
When, for $\sigma >1$, $z$ takes the same value $z=2$ as in the case with nn interactions, the motion of interfaces
behaves as for nn interactions, namely it is diffusive for $d=1$ and is governed by the 
curvature~\cite{BrayRut94,RutBray94} for $d>1$.
Instead, when $z=z_{\rm adv}$, i.e. for $\sigma \le 1$, the motion 
of domains walls is advected
by the drift due to the long-range interactions between far away spins. Notice that
one has $z\to z_{ball}=1$ for $\sigma \to 0^+$. This case corresponds to an advection of interfaces so strong 
to produce a completely deterministic motion and hence a ballistic regime. This is related 
to the crossover from weak long-range to strong long-range occurring right at $\sigma =0$.

Regarding the differences between $T>0$ and $T=0$ when long-range interactions are present, the 
situation is not clear at all. Presently the matter has been well understood only in 
$d=1$~\cite{CLP_review,CLP_lambda,PhysRevE.102.020102}.
In this case, when quenching to $T=0$ one finds coarsening with $z=z_{ball}=1$ for any $\sigma$. Let
us recall that, instead, for $T>0$ one has $z\to 1$ only in the limit
$\sigma \to 0^+$. The interpretation is the following: when $T>0$ thermal fluctuation randomise the displacement
of interfaces and, therefore, the motion is not fully deterministic although still advected. In this case, in order
to have full determinism,  one has to go to the strong interaction limit $\sigma \to 0$. Instead, when $T=0$,
even a relatively small drift, present for any $\sigma \le 1$, leads to a deterministic motion with $z=z_{ball}=1$. 
Also in this case, the asymptotic ballistic growth which sets in at $T=0$ is observed
as a pre-asymptotic behaviour after quenches to finite $T$ for any $\sigma $.
Finally, let us mention that also with long-range interactions, as in the nn case, metastability is not observed
in $d=1$.   

In this paper we take a first step in the direction of understanding the ordering kinetics after quenches to 
$T=0$ of systems with long-range interactions in $d>1$. The matter, which is largely unexplored, is relevant because in this case, as we shall 
illustrate, the dynamical mechanism is different from the ones discussed above, 
thus producing a new value $z={\cal Z}$ of the growth exponent. This is independent of $\sigma$ and 
characterises also the  
pre-asymptotic evolution in deep quenches to a finite $T$. 

The origin of this new growth mechanism can be traced back to metastable configurations. 
In order to discuss this point we invite the reader to refer to Fig.~\ref{stab_interf} as a preliminary illustration of the
various shapes of the interfaces, leaving further details contained in this figure, described in the detailed caption,
to the specific discussion that will be conducted in Sec.~\ref{sec.stab}.
As discussed previously, in $d>1$ with nn interactions a flat portion of an interface along a lattice direction 
(Fig.~\ref{stab_interf}, letter $a$)
is stable at $T=0$.
However, interfaces are never perfectly flat neither aligned along the lattice directions in the coarsening stage, and hence they can be deformed 
by flipping spins on the edges (Fig.~\ref{stab_interf}, letters $b,c,d$). This process typically occurs without any drift and 
hence has a diffusive character~\cite{CorLipZan08}, a fact that leads to an exponent $z=z_{\rm diff}=2$.
The addition of long-range interactions changes the situation in two fundamental respects. The first modification 
is an increase
of (meta)stable configurations: any globally straight interface (e.g. the one of Fig.~\ref{stab_interf}, letter $b$, where
all steps have the same size)
is now locally stable, even if its direction does not fit the orientation
of the underlying lattice. 
This means that also spins on edges can be blocked, at variance with the nn case. 
Secondly, if we perturb the constant slope interface (e.g. see Fig.~\ref{stab_interf}, letters $c,d$, where steps 
have different lengths), 
spins which are now free to flip are subjected to a deterministic drift, similarly to what happens in $d=1$
in the ballistic regime. These two new ingredients have, so to say, contrasting effects. On the one hand the presence 
of the drift tends to speed the kinetics with respect to a diffusive case, leading to $z<2$;
on the other hand the drift is contrasted by the abundance of blocked states, causing $z>1$. 
The result is a non trivial value, $z={\cal Z}$, with $1<{\cal Z}<2$. Even if this regime is observed unambiguously 
in numerical simulation, a precise determination of ${\cal Z}$ is difficult with this technique also because
a pre-asymptotic stage is present. A direct analytical attack of the problem 
seems to be difficult as well. This is due to the presence of long-range correlations which mix up with 
lattice effects making calculations elusive. 
Indeed, as we will discuss further, an approach on the continuum, i.e. off-lattice, which accurately
captures the asymptotic values $z_{\rm adv}, z_{\rm diff}$ in quenches to finite $T$, provides 
a wrong value of ${\cal Z}$. 
Considering instead a simplified system with a single domain, we are able to obtain a full analytic understanding
of the early stage where ${\cal Z}=3/2$ and a quite clean numerical determination of its asymptotic one with ${\cal Z}=4/3$. While the former value can be traced back to a relatively simple circular geometry of the growing domains, the latter originates from a more peculiar one caused by the
long-range interactions extending along the interfaces.

This paper is organised as follows: In Sec.~\ref{model} we introduce the kinetic model we study, discuss the
implementation of the numerical simulations and define the basic observable quantities we compute.
Sec.~\ref{numer_res} presents the results of numerical simulations of the model and the determination
of the exponent ${\cal Z}$. In Sec.~\ref{simplified} we simplify the problem by considering the evolution
of a single domain. This is studied both numerically and analytically, allowing us to 
determine ${\cal Z}$. Finally, in Sec.~\ref{concl}, we conclude the paper by summarising the main results and 
discuss some open points. In Appendix~\ref{Ewald}, we give some details of the numerical technique, namely Ewald summation, to incorporate long-range interactions in the system.

\section{The model and its numerical simulation}
\label{model}

We consider an Ising model with the Hamiltonian
\be
{\cal H}(\{s_i\})=-\frac{1}{2} \sum _i \sum_j{}^{'} J(r) s_i s_j,
\label{ham}
\ee
where $s_i=\pm 1$ are boolean spin variables on the sites $i$ of a two-dimensional square lattice
of liner size ${\cal L}$, whose spacing we assume to be unitary. The sum over $j$ is primed to indicate that the terms with $i = j$ are excluded, and
\be
J(r)=r^{-(2+\sigma)}
\ee
is a ferromagnetic coupling constant with $r=\vert \vec r_i - \vec r_j \vert$; $\vec r_{i}$ being the coordinate of site $i$ of the lattice. The usual nn Ising model arises on setting $J(r)=\delta _{r,1}$.

At equilibrium the long-range model has a para-ferromagnetic phase transition at a finite critical temperature $T_c(\sigma)$.
Critical exponents 
match~\cite{PhysRevB.1.2265,Fisher1972critical,PhysRevB.8.281,PhysRevB.56.8945,PhysRevLett.89.025703,PhysRevE.95.012143} those of the corresponding nn model for sufficiently large values of $\sigma$,
i.e. $\sigma \ge \sigma_{nn}$, where $\sigma _{nn}$ has been estimated~\cite{PhysRevB.8.281,PhysRevE.95.012143} to be $\sigma_{nn}=7/4$,
whereas they match the mean field values for $\sigma <1$. In between, for $1< \sigma < \sigma_{nn}$, critical 
exponents depend continuously on $\sigma$.

A kinetics is introduced by flipping single spins with Metropolis transition rates
\be
w(s_i\to -s_i)={\cal L}^{-2}\min \left\{1,e^{-\frac{\Delta E}{T}}\right \},
\label{metrop}
\ee
where $\Delta E$ is the change in energy due to the spin flip to be attempted, and we have set to unity
the Boltzmann constant. Time is measured in Monte Carlo steps (MCS), each of which corresponds to
${\cal L}^2$ elementary spin-flip attempts.
We consider a quenching protocol where the system is initialised in a configuration sorted from an infinite 
temperature equilibrium ensemble, i.e. spins are randomly and independently set to $+1$ or $-1$.
This initial state is evolved at the quench temperature $T$ by means of the transition probabilities~(\ref{metrop}) 
with $T<T_c(\sigma)$.

Since the spin-spin interaction is long-ranged, the calculation of $\Delta E$ at every spin-flip attempt via Metropolis probability~(\ref{metrop}) is computationally expensive. To speed up the computer simulation, we store the local field for each spin at the beginning of the simulation. In this way we need to update it only once the spin-flip is accepted. This simple trick significantly speeds up the numerical calculations at any given quench temperature \cite{hmu1995,CMJ19}.

The main issue during the simulations of the system with Hamiltonian~(\ref{ham}) is the strong finite-size effect arising due to the long-range character of $J(r)$. One obvious way to diminish these effects is to use periodic boundary conditions via minimum-image convention \cite{fs2002}. In this approach, the square lattice is mapped onto a torus where each spin in the system interacts with other spins up to a certain cut-off distance ($\le {\cal L}/2$). When the interaction decays slowly, this natural cut-off limit on the interaction-range causes artefacts in the simulation results. Therefore, we need a more sophisticated approach to implement periodic boundary conditions in such systems. For this, we envision an infinite $2d$-lattice partitioned into infinite imaginary copies of the original simulation lattice. The central cell of this infinite lattice is the simulation lattice itself, and the imaginary copies, called images, lie across its periodic boundaries in both $x$ and $y$-direction. Implementing periodic boundary conditions with infinite images remove cut-off errors in the simulation results of long-range interacting systems. The effective interaction between two spins inside the simulation lattice can now be expressed as an infinite summation over all images
\begin{equation}
\label{jsum}
J(s_i,s_j) = \sum_{n_x} \sum_{n_y} \frac{1}{\vert \vec n + \vec r_{i} - \vec r_{j} \vert ^{2+\sigma}},
\end{equation}
where displacement vector $\vec n = (n_x,n_y)$ with $n_x, n_y = 0, \pm {\cal L}, \pm 2{\cal L}, ...$, representing the coordinates of image systems and the simulation lattice is located at $\vec n = (0,0)$. The infinite summation involved in (\ref{jsum}) has slow convergence in coordinate space; therefore, it is difficult to handle it directly during the numerical simulations. We have adapted the Ewald summation technique \cite{fs2002,ew1921,PhysRevE.95.012143}, which uses a clever trick to split it into two independent rapidly convergent summations, one in coordinate space and another in reciprocal space (see Appendix~\ref{Ewald}).

The main observable we consider in this paper is the characteristic size of the growing domains,
which we extract from the time dependent spin configurations by means of the equal time
correlation function
\be
C(r,t)=\langle s_i(t)s_{i+\vec r}(t)\rangle,
\label{defcor}
\ee
where $\langle \dots \rangle$ is an off-equilibrium average, namely taken over different initial conditions
and thermal histories. If dynamical scaling holds, in quenches below $T_c$ this quantity depends
on a single variable as~\cite{PuriWad09,Bray94,CRP,Cor15,CorCugYos11}
\be
C(r,t)=f\left (\frac{r}{L(t)}\right ),
\label{cscal}
\ee
where $f$ is a scaling function and $L(t)$ has the meaning of the size of the growing domains at time $t$.
This quantity can be extracted from the correlation function itself as \cite{PhysRevE.75.011113,PhysRevE.74.041106,CMJ19}
\be
C\left (r=L(t),t\right )= \frac{C(0,t)}{2} \equiv \frac{1}{2} .
\ee

The growth law after a quench to $T>0$ was predicted in~\cite{BrayRut94} by Bray and Rutenberg to be characterised by $z=z_{\rm cd}=2$ for $\sigma >1$ and by $z=z_{\rm adv}=1+\sigma$
for $\sigma \le 1$, with logarithmic corrections at $\sigma =1$. Such a prediction was obtained by using a 
continuum model, based on a Ginzburg-Landau free energy, assuming a dynamical scaling symmetry and 
resorting to a an energy scaling argument. These results have  been recently confirmed by Christiansen et al. \cite{CMJ19} by means of numerical simulations of the model considered in the present paper.

In order to study the growth law $L(t)$ it is useful to consider the effective exponent $z_{\rm eff}(t)$
defined by $1/z_{\rm eff}(t)=d\ln L(t)/d\ln t$. 
Since, in order to speed our simulations, observable quantities are computed
only at discrete times $t_i$ (equally spaced in $\ln t$), the effective exponent is computed as follows 
\begin{equation}
1/z_{\rm eff}(t_i) = \frac{\ln L(t_i)-\ln L (t_{i-1})}{\ln t_i-\ln t_{i-1}}.
\label{eqeffexp}
\end{equation}
The effective exponent computed by means of the definition~(\ref{eqeffexp}) may have 
a noisy character, therefore we will also use the determination based on the symmetric derivative, which
amounts to 
\begin{equation}
1/z_{\rm eff}^{(s)}(t_i)=\frac{ \ln L(t_{i+1})- \ln L(t_{i-1}) }{ \ln(t_{i+1}) - \ln(t_{i-1}) }.
\label{eqeffexps}
\end{equation}

\section{Numerical results}
\label{numer_res}
  
\begin{figure}[h]
	\vspace{1.5cm}
	\centering
	\rotatebox{0}{\resizebox{.99\textwidth}{!}{\includegraphics{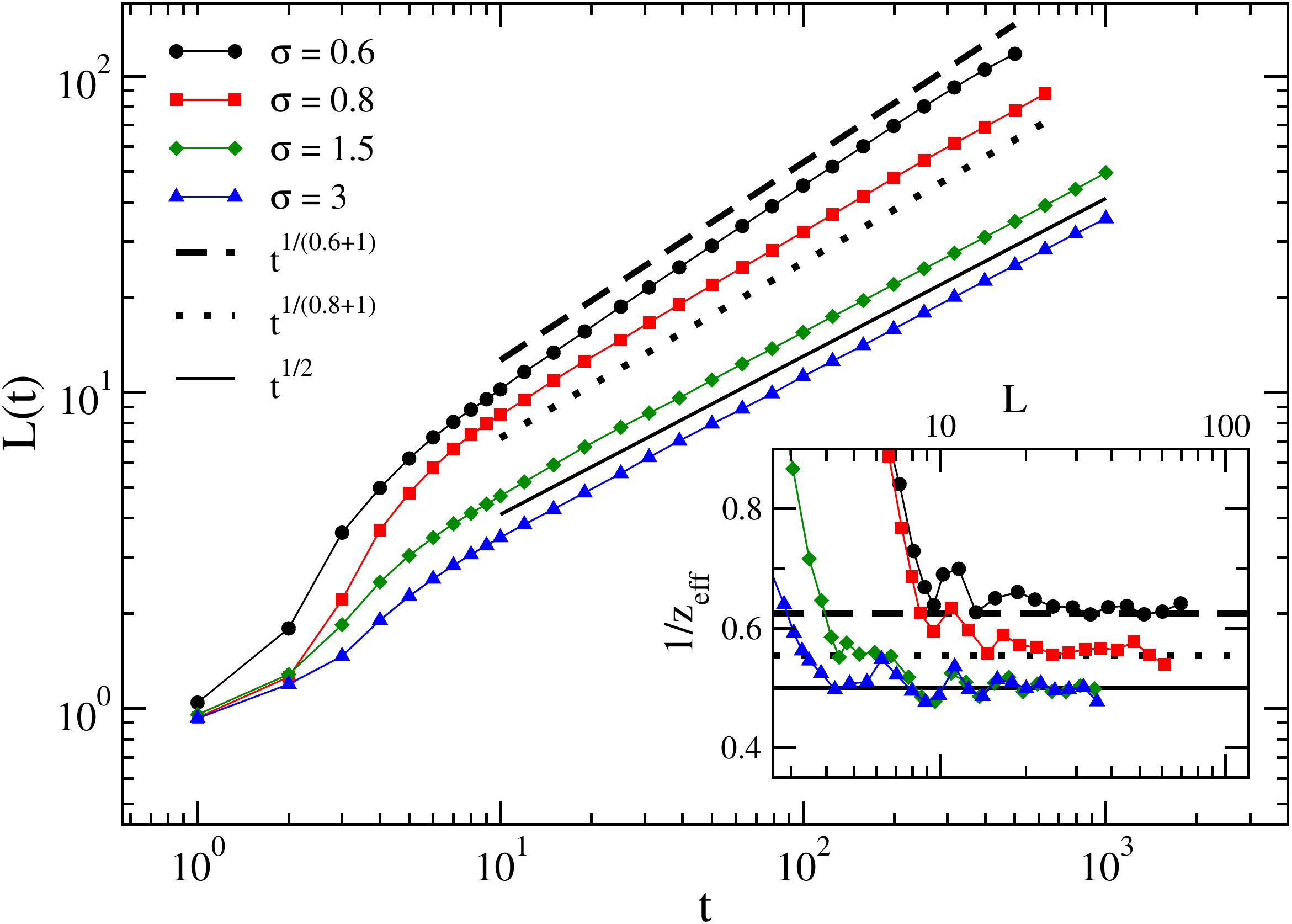}}}
	\caption{
		$L(t)$ is plotted against $t$ on a double logarithmic scale for systems with various values of $\sigma$ (see key)
		quenched from infinite temperature to final temperatures $T=1.255$, $T=2.929$, $T=1.5$, $T=0.9$ for 
		$\sigma=0.6,0.8,1.5,3$, respectively (such temperatures are in the range $0.1 T_c - 0.3 T_c$).
		The system size is $2048^2$ and each data set is averaged over 10 realisations. 
		The different lines (see keys) represent the expected asymptotic behaviour
		$L(t)\sim t^{1/z}$ with $z=1+\sigma$ for $\sigma \le1$ and $z=2$ for $\sigma >1$. In the inset
		the effective exponent $1/z_{\rm eff}$ is plotted against $L(t)$ on a log-linear scale. The horizontal dashed, dotted and solid lines are the 
		expected asymptotic values.}
	\label{asymptoticRB}
\end{figure}

Let us now discuss the outcomes of our numerical simulations. First,  as a benchmark, we show
in Fig.~\ref{asymptoticRB} the results for quenches to relatively high $T$, where we expect 
to recover similar results to those found by Christiansen et al.~\cite{CMJ19}, thus confirming the 
Bray-Rutenberg asymptotic exponents. In this figure we see that an algebraic growth of $L(t)$
sets in at times $t\gtrsim 10$, for all $\sigma$ values. The value of $z$ matches with
the prediction of Bray and Rutenberg, as it can be appreciated in the main figure and in the inset,
where the effective exponent $1/z_{\rm eff}$ is plotted against $L(t)$. 

\begin{figure}[h]
\vspace{1.5cm}
  \centering
  \rotatebox{0}{\resizebox{.49\textwidth}{!}{\includegraphics{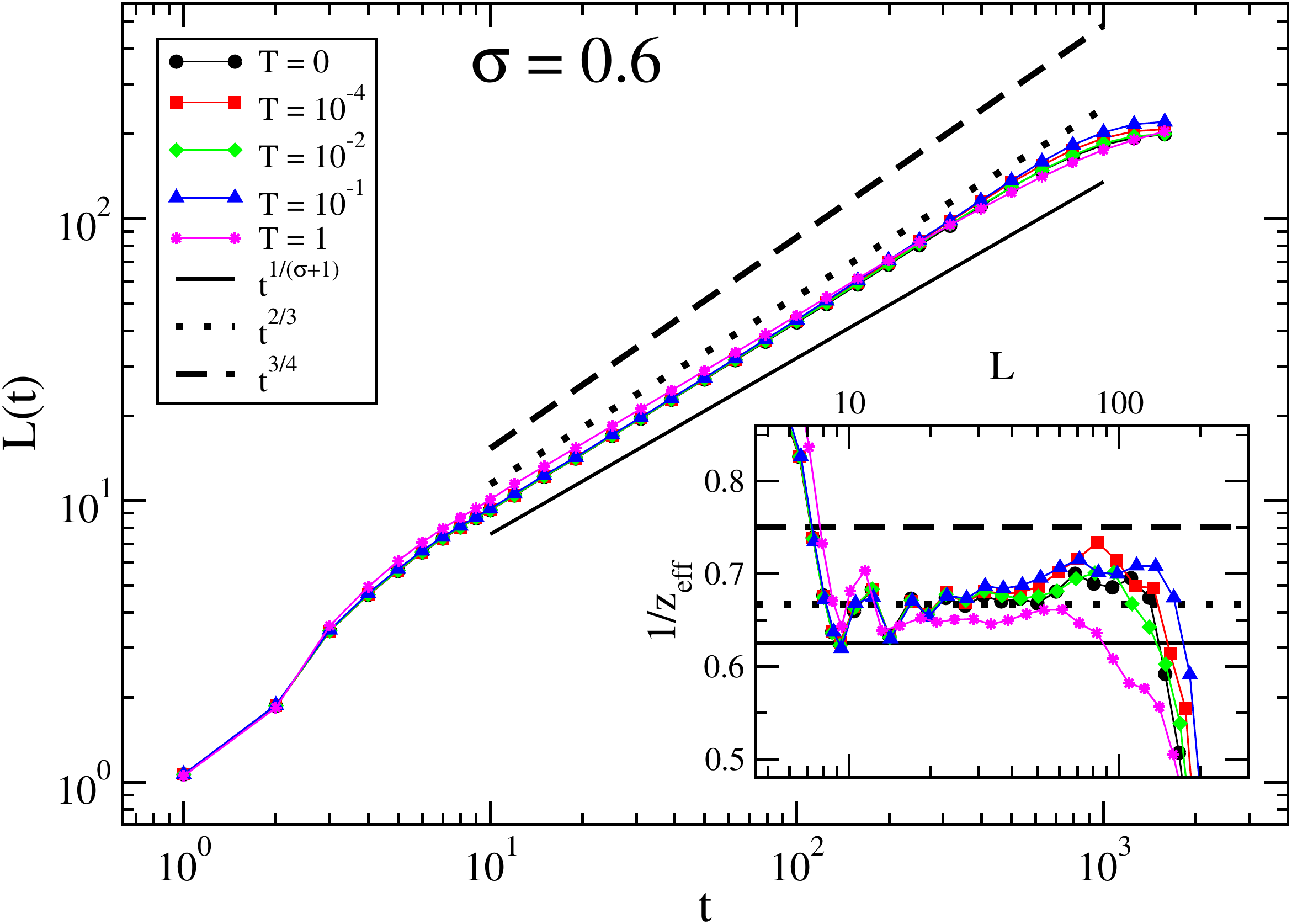}}}
    \rotatebox{0}{\resizebox{.49\textwidth}{!}{\includegraphics{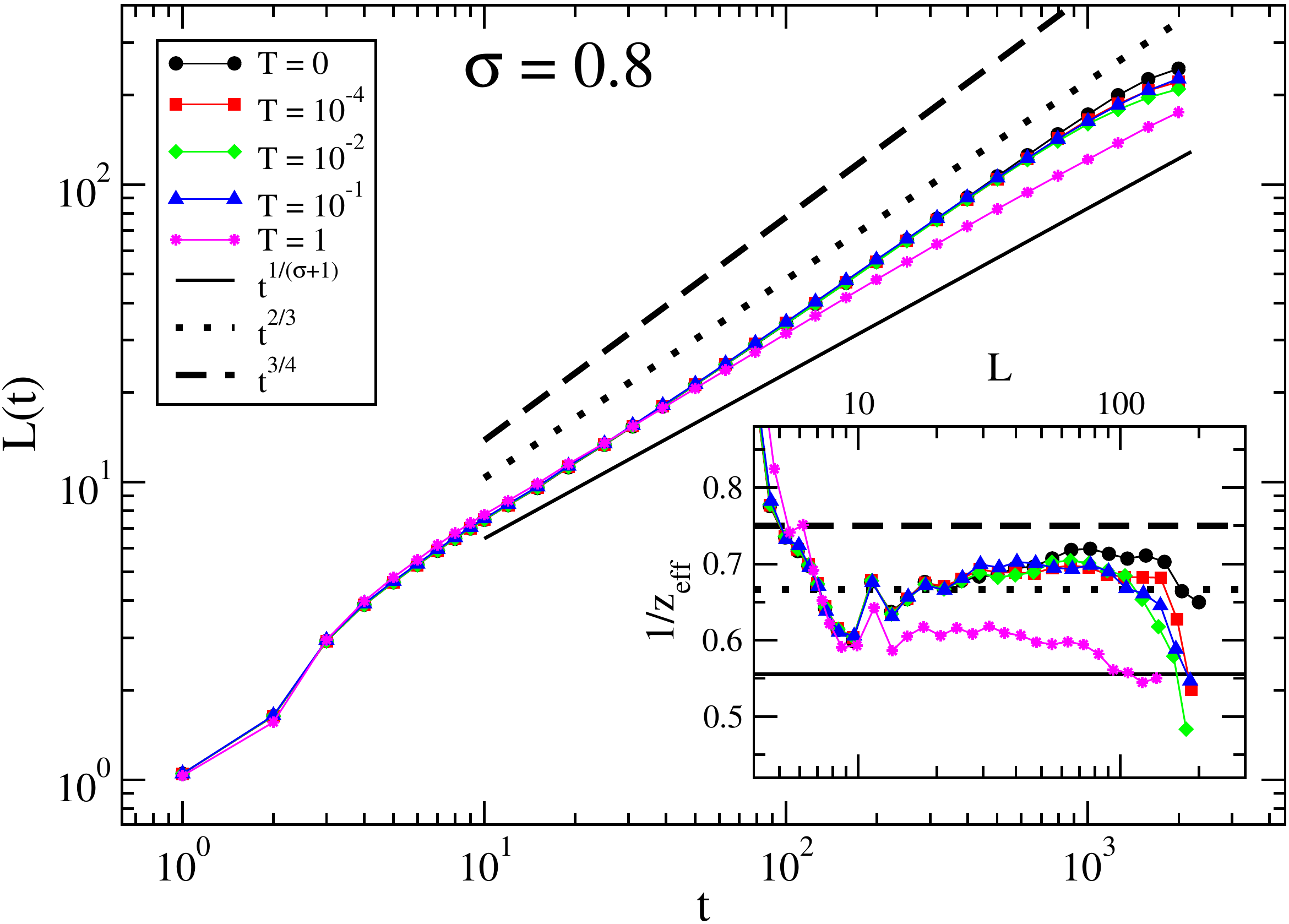}}}
      \rotatebox{0}{\resizebox{.49\textwidth}{!}{\includegraphics{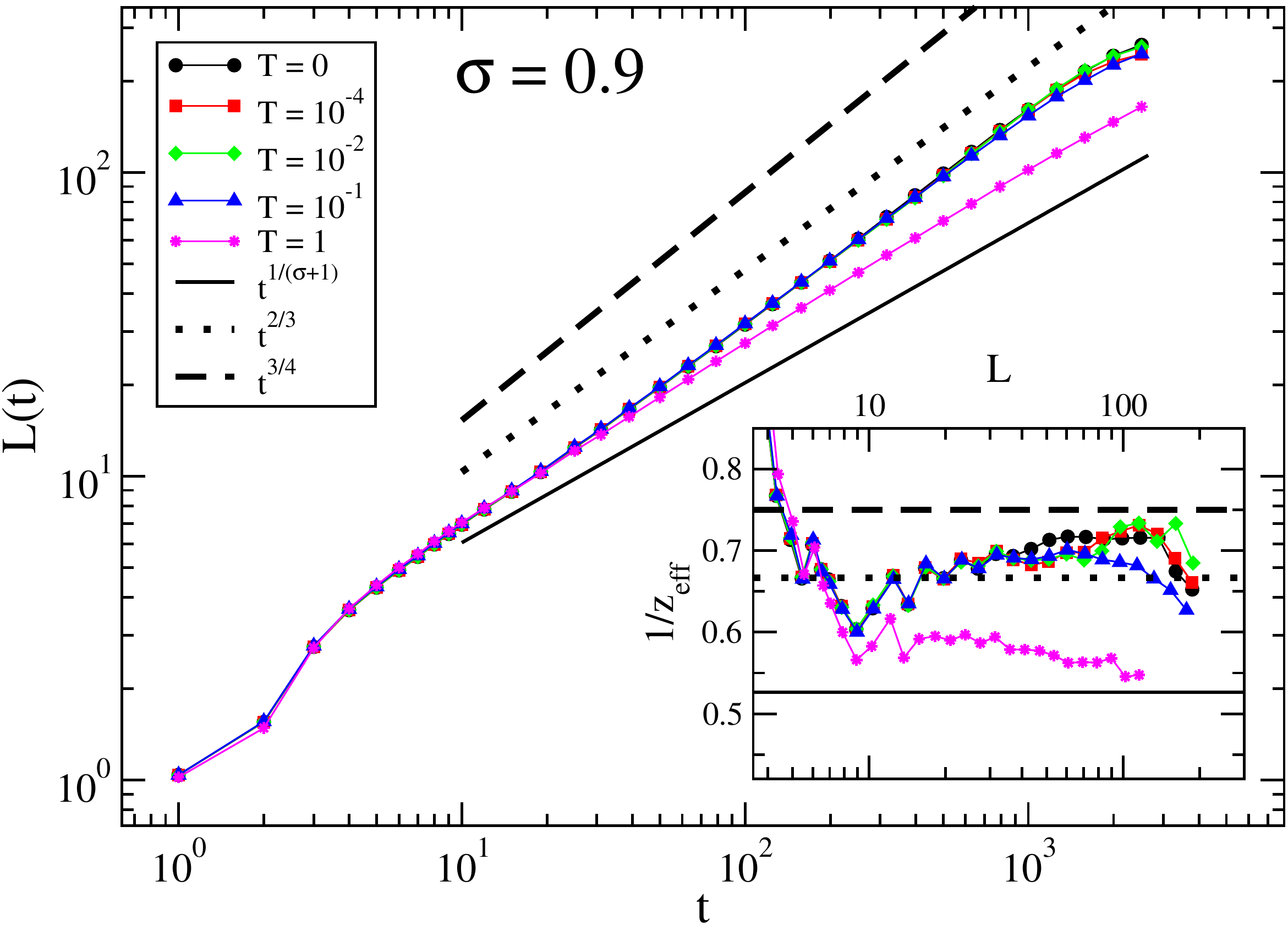}}}
        \rotatebox{0}{\resizebox{.49\textwidth}{!}{\includegraphics{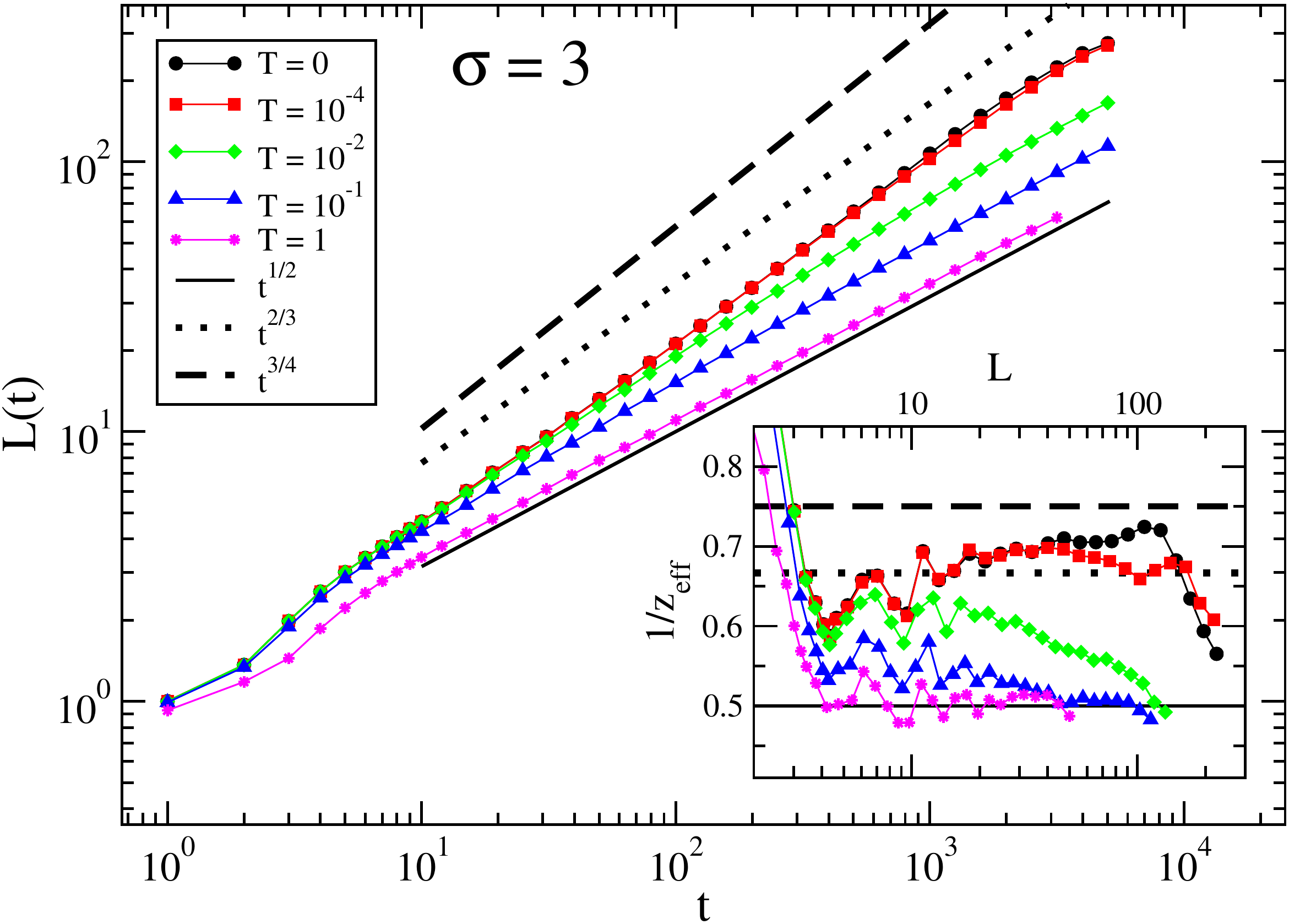}}}
  \caption{
$L(t)$ is plotted against $t$ on a double logarithmic scale for systems with various values of $\sigma$ (see keys)
  quenched from infinite temperature to different final temperatures (see keys).
  The system size is $2048^2$ and each data set is averaged over 40 realisations. 
  The solid, dotted and dashed lines (see keys) represent the expected asymptotic behaviour at finite temperatures and the two putative
  behaviours $t^{2/3}$ and $t^{3/4}$ expected pre-asymptotically and at late times at zero and low temperatures.  
  In the inset the effective exponent $1/z_{\rm eff}$ is plotted against $L(t)$ on a log-linear scale. The horizontal solid, dashed and dotted lines are the 
  above discussed expected asymptotic and pre-asymptotic exponents.}
\label{preasympt}
\end{figure}

Let us now move to the core results of this article, namely the effect of quenching to $T=0$. 
In Fig.~\ref{preasympt} we summarise the behaviour of $L(t)$ not only in zero temperature quenches,
but also in deep quenches to small but finite $T$, in order to appreciate the existence of pre-asymptotic
effects. Let us start to discuss the case with $\sigma =3$ (lower right panel). 
Since this value of $\sigma $ is rather large one could
naively expect to see a behaviour akin to that of the nn model. As we know, such a naive argument is correct
when applied to the shallow quenches discussed above, since for any $\sigma >1$ one recovers the 
exponent $z_{\rm cd}=2$ of the nn case. Instead here one observe that for $T=0$ (black curve), for
$t\gtrsim 10$ one has an algebraic increase of $L(t)$ with an exponent $1/z\simeq 0.7$, definitely
larger than the one, $1/z_{\rm diff}=0.5$, of the corresponding quench to $T=0$ in the model with nn interactions
(and of the one $1/z_{\rm cd}=0.5$ predicted by Bray-Rutenberg for quenches to $T>0$ with $\sigma >1$).
Although a precise determination of $z$ is not possible from these simulations, this is surely
enough to establish the existence of a new exponent, that we denote with ${\cal Z}$, associated to the
zero temperature quenches with long  range interactions. 
Notice the decrease of such exponent
at large $L(t)$ (meaning very large times) must be attributed to finite size effects start to be
appreciated. 

The same panel of Fig.~\ref{preasympt} shows that a quench to 
$T=10^{-4}$ behaves very similarly to the case with $T=0$, whereas a quench to $T=10^{-2}$
does so only up to times smaller than $t\simeq 10^2$, after which $L(t)$ slows gradually down until being
compatible with $1/z_{\rm eff}=1/z_{\rm cd}=1/2$. This pattern of behaviours can be interpreted as a crossover
occurring at $L_{\rm cross}(T,\sigma)$ between an early regime with $z={\cal Z}$ and a late stage with the
Bray-Rutenberg exponent. Since $L_{\rm cross}(T,\sigma)$ is a decreasing function of $T$ the crossover
cannot be observed in the quench to $T=10^{-4}$ because it occurs after the longest simulated times.

\begin{figure}[h]
\vspace{1.5cm}
  \centering
  \rotatebox{0}{\resizebox{.8\textwidth}{!}{\includegraphics{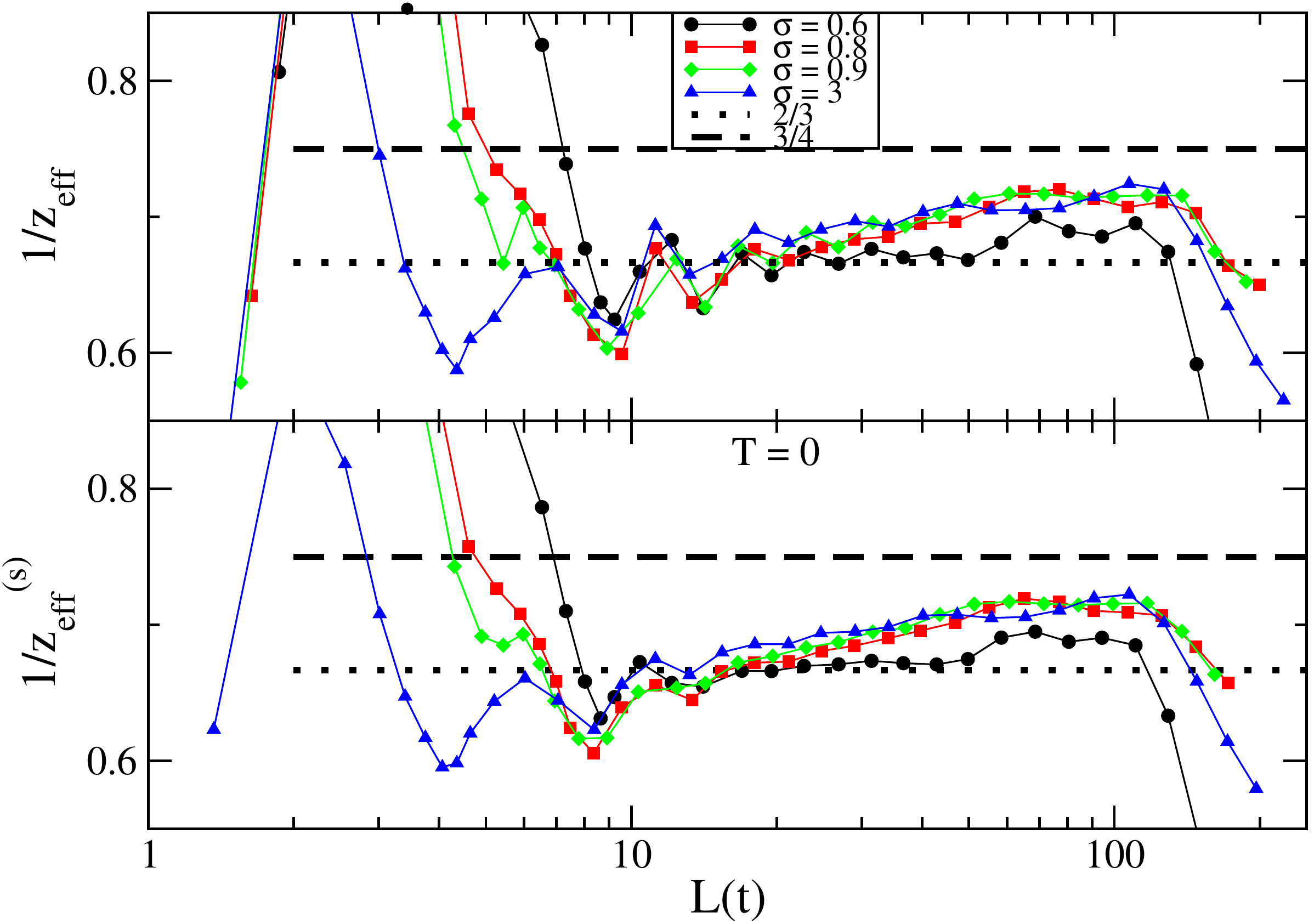}}}
    \rotatebox{0}{\resizebox{.8\textwidth}{!}{\includegraphics{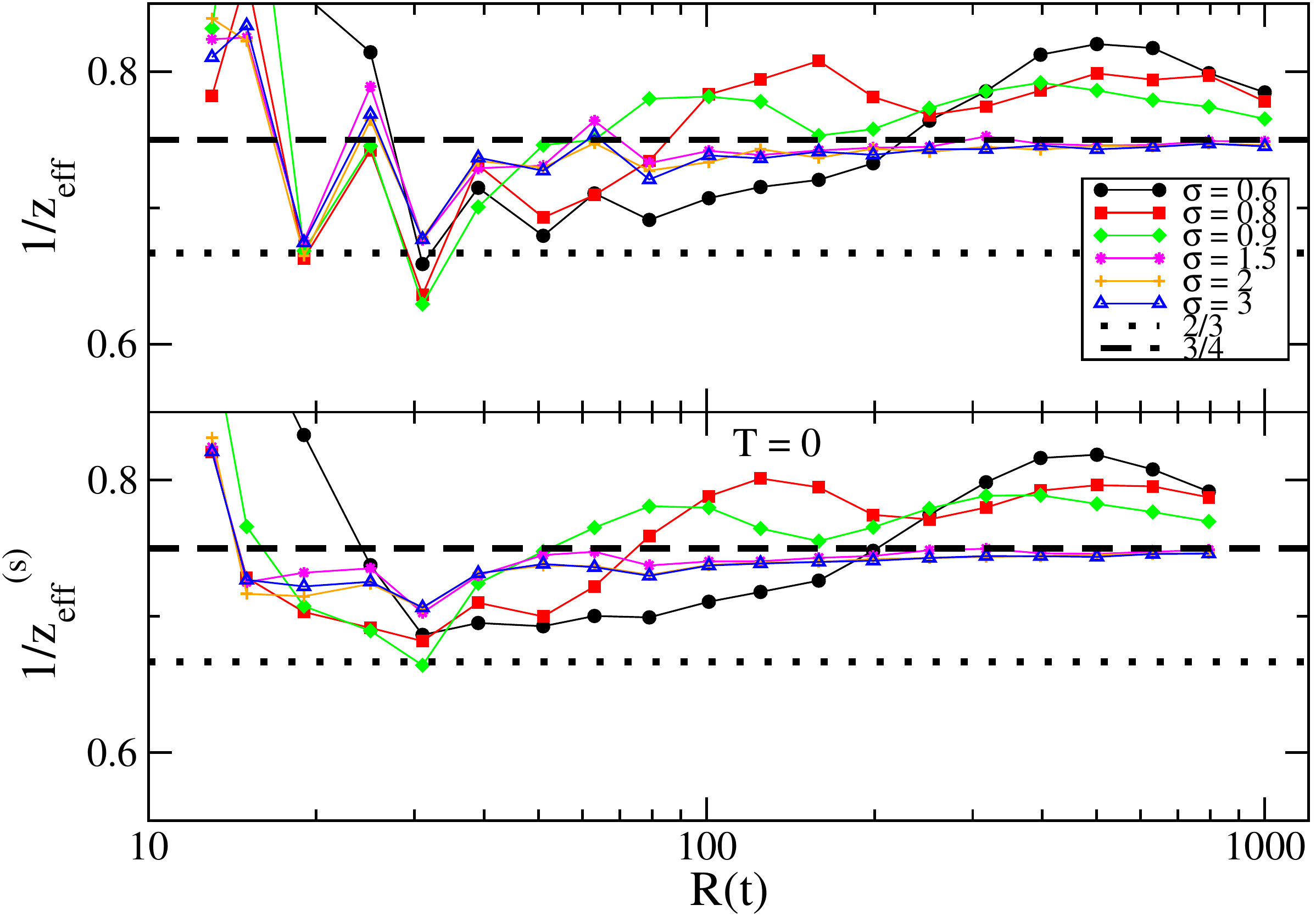}}}
  \caption{
A comparison between the effective exponent $1/z_{\rm eff}$ and $1/z_{\rm eff}^{(s)}$ in the quench to $T=0$ with various
values of $\sigma$ is presented on a log-linear scale. The top panel contains the same data as in Fig.~\ref{preasympt} for the coarsening multi-domain system (only for $T=0$). The lower panel contains the same data as in 
Fig.~\ref{bubble} for the single shrinking bubble model. 
The dashed line and the dotted one are the values $3/4$ and $2/3$.}
\label{summary_exp}
\end{figure}

Moving now to the other values of $\sigma $ considered in Fig.~\ref{preasympt} we observe a similar
pattern of $L(t)$, with a regime characterised by a value
of $1/z$ of order $0.7$. This value, with the precision of our numerical simulations, appears to be 
roughly independent of $\sigma$, a fact that suggests that ${\cal Z}$ is a universal exponent. 
This can be better appreciated in the top panel of Fig.~\ref{summary_exp}, where we compare $1/z_{\rm eff}$ 
and $1/z_{\rm eff}^{(s)}$ in zero
temperature quenches with various $\sigma$. This figure shows that $1/z_{\rm eff}$ sets to a value of order
$1/z_{\rm eff}\simeq 2/3$ for $L(t)$ around $L(t)=10$, and then slowly increases towards a value that 
exceeds $0.7$, before bending down due to finite size effects. A similar behaviour,
somewhat less noisy, is displayed by $1/z_{\rm eff}^{(s)}$. This observation, together
with the study of single domain models that we will discuss in Section~\ref{simplified}, leads us to the
conjecture that the exponent ${\cal Z}$ toggles between a pre-asymptotic value ${\cal Z}=3/2$ and an
asymptotic one ${\cal Z}=4/3$. These two values are indicated by straight lines in Figs.~\ref{preasympt}
and \ref{summary_exp}.

Comparing the various panels of Fig.~\ref{preasympt} 
one can also be convinced that the crossover length $L_{\rm cross}(T,\sigma)$ is a (rather strongly) decreasing
function of $\sigma$. Indeed, for instance, with $\sigma =0.9$ (lower left panel) one has to rise $T$
by a factor $10^2$ (i.e. to set $T=1$) in order to see a crossover pattern similar to the one observed
at $\sigma =3$ (at $T=10^{-2}$).
Moreover, one observes that finite-size effects are more important for the smaller values of $\sigma $ 
because the longest range of the interactions makes the system feels the periodic boundary conditions  earlier.

\begin{figure}[h]
\vspace{1.5cm}
  \centering
  \rotatebox{0}{\resizebox{.99\textwidth}{!}{\includegraphics{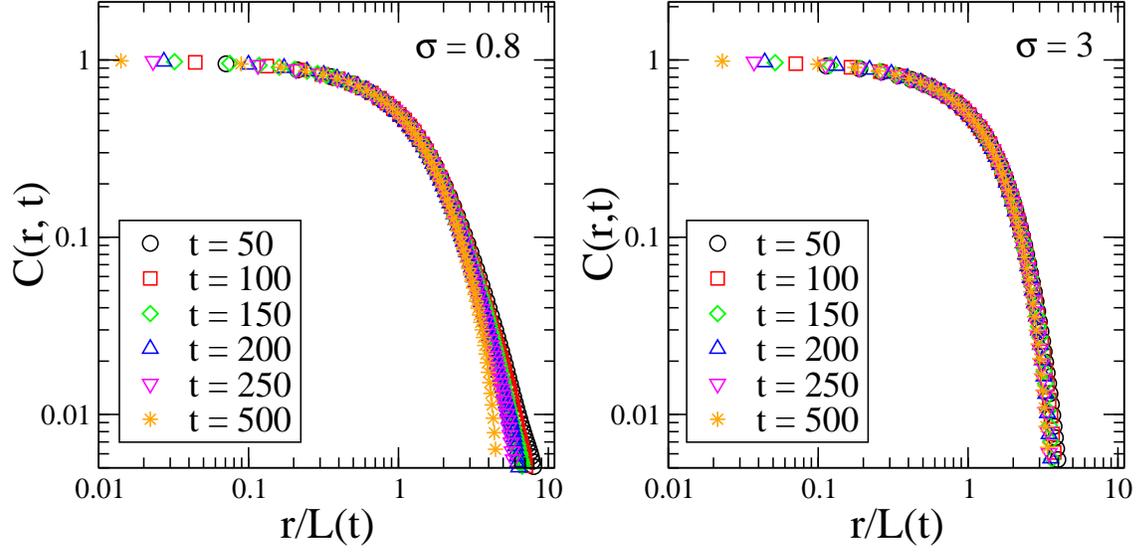}}}
  \caption{$C(r,t)$ is plotted against $r/L(t)$
on a double logarithmic scale for systems with $\sigma=0.8$ (left panel) and $\sigma=3$ (right panel)
  quenched from infinite to zero temperature. Different symbols and colours refer to different times,
  as in the key.  
  The system size is $2048^2$ and each data set is averaged over 40 realisations.}
\label{corr_log}
\end{figure}

Finally, let us discuss the issue of dynamical scaling. This was shown to be obeyed~\cite{CMJ19} in the 
Bray-Rutenberg regime occurring in quenches to a finite $T$, by checking the form~(\ref{cscal}) of
the correlation function. In the following we study the same matter in the zero-temperature
quenches. Our results for $C(r,t)$ are shown in Fig.~\ref{corr_log}. The data for different times show
a very good collapse when plotted against the rescaled variable $r/L(t)$, as expected after Eq.~(\ref{cscal}) in
the presence of dynamical symmetry. This is true for all the values of $\sigma $ considered (we show only
a couple of them in Fig.~\ref{corr_log} but similar results are found for the other values). 
For $\sigma =0.8$ (left panel), the curve at the longest time $t=500$ starts departing from the 
collapse, but this is due to the onset of finite-size effects which can also be detected at these times
from the behaviour of $L(t)$ shown in Fig.~\ref{preasympt}. Hence, we conclude that the dynamical scaling
symmetry is at work also in zero temperature quenches, a fact that will be used in the following. Let us also 
mention that metastable states, which are very important with nn interactions, here are absent (or at least greatly
suppressed), as found also in~\cite{Christiansen_arxiv}.   

\section{Evolution of a single domain}
\label{simplified}

In the previous section we have shown that, with long-range interactions, one observes a fast growth
regime at $T=0$ regulated by a new exponent $1/{\cal Z}\simeq 0.7$. However, in the absence of
analytic approaches or of a detailed comprehension of the mechanism at work a precise determination
of such exponent on the basis of the sole simulations could not be obtained. 

\begin{figure}[h]
  \centering
  \rotatebox{0}{\resizebox{.7\textwidth}{!}{\includegraphics{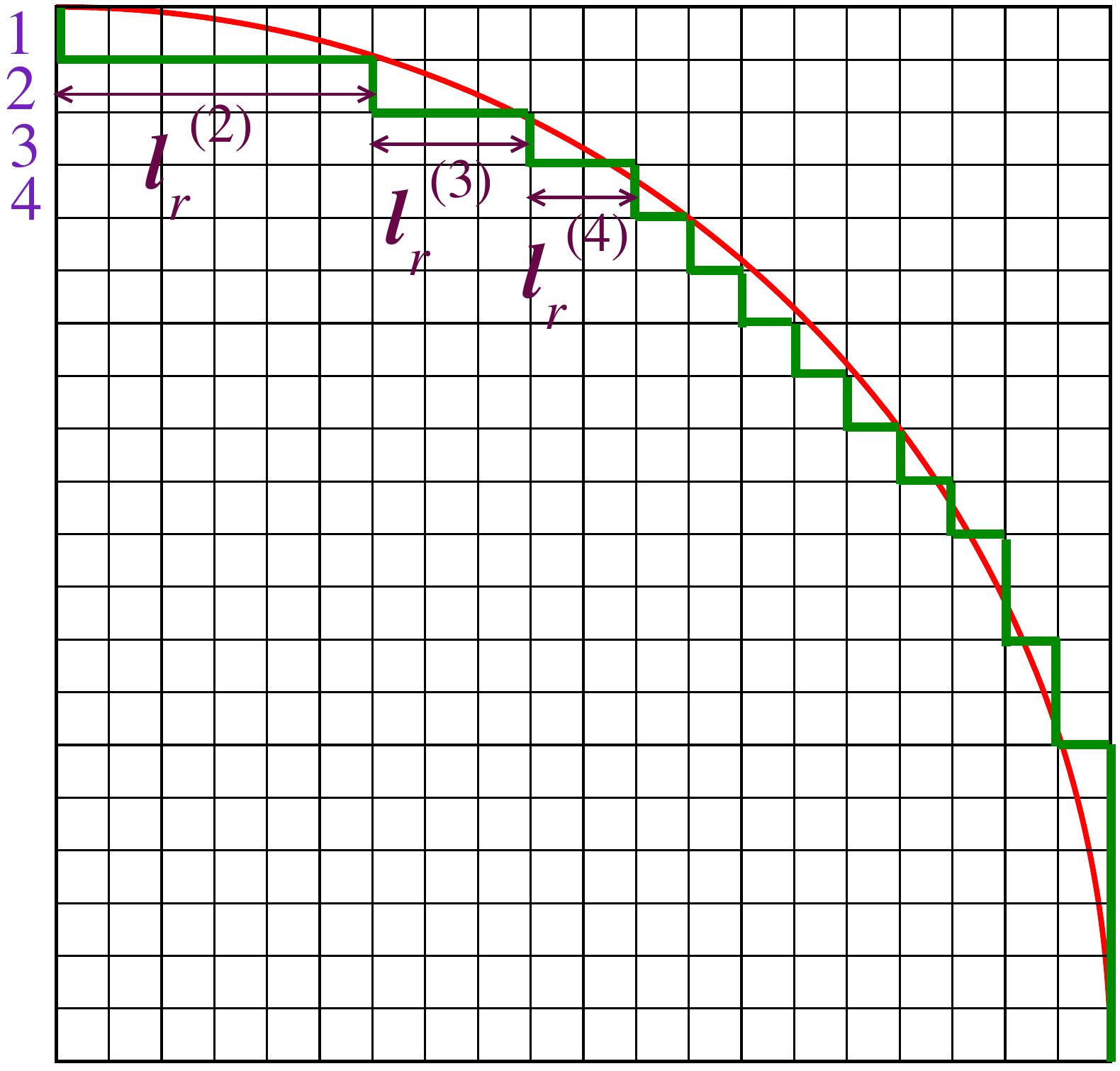}}}
  \caption{Pictorial representation of a circular domain (only one quadrant) on a square lattice. 
  Spins are located in the centre of 
  the small squares forming the lattice. The domain can be viewed as the superposition of terraces, numbered according
  to the violet figures near the vertical axis. The droplet is shown at a time when the upper terrace (number 1) has shrunk to 
  zero and is disappearing. In this configuration the length of the $j$-th terrace is indicated with ${\ell}_r^{(j)}$. 
  With the simplified dynamics described in Sec.~\ref{Z_shape} only the edge spins can flip. The red curve is a circle in the continuum.}
\label{pict_domain}
\end{figure}

\begin{figure}[h]
\vspace{1.5cm}
  \centering
  \rotatebox{0}{\resizebox{.8\textwidth}{!}{\includegraphics{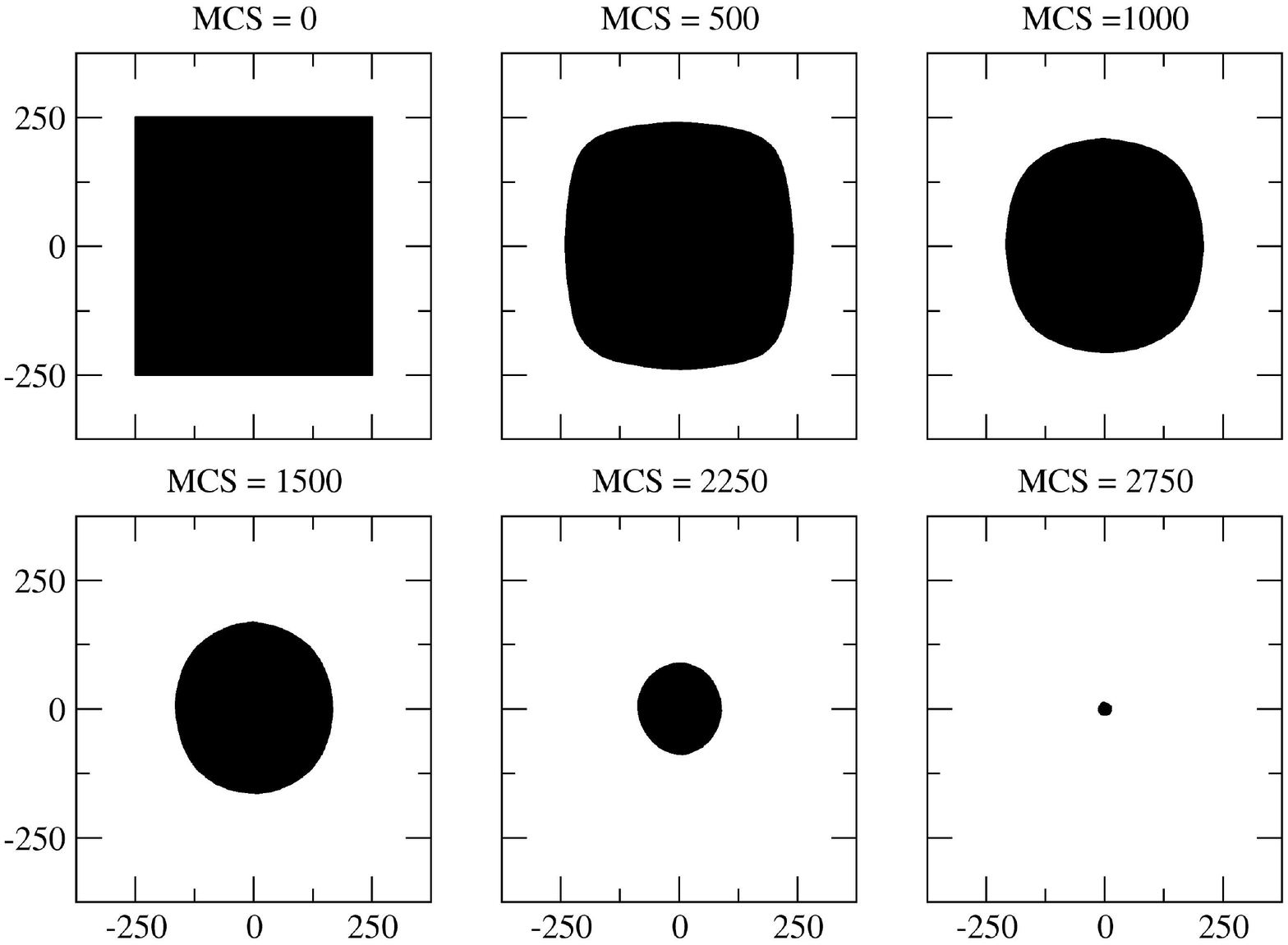}}}
  \caption{The shrinkage of an initially square bubble of size 500 is shown for $T=0$ and $\sigma=3$.
  Different snapshots correspond to the simulation times reported on the top of the panels (times are
  expressed in Monte Carlo steps).}
\label{bubble_geom}
\end{figure}

In this section we study the simpler process with a single domain. These approach has been shown
to provide a correct determination of the growth exponent $z$ in systems with nn interactions, 
for any $d$~\cite{CorLipZan08,CLP_epl,libro}, and also with long-range interactions in $d=1$~\cite{CLP_review}. 
We consider the same model of Sec.~\ref{model} with the only difference that 
the initial state is not sorted from an infinite temperature ensemble but is built by hand
with a single domain of size $R$. This configuration is evolved with the
flip probabilities~(\ref{metrop}), which lead to a shrinkage and eventual disappearance
of the bubble. We start with a circular shape of radius $R$ (more precisely,
the lattice approximation to a circle, as in Fig.~\ref{pict_domain}), because we observed that, 
starting from different shapes, a roughly circular one tends to be formed during the evolution
after a transient (however, the true asymptotic shape is not perfectly circular, see discussion in Sec.~\ref{Z_shape}). 
This is shown in Fig.~\ref{bubble_geom} for an initially square bubble. Notice that the late
geometry of the bubble almost isotropic, at variance with what observed with nn interaction
during coarsening~\cite{PhysRevE.54.R2181} or in metastable states~\cite{GUNTHER1994194}.
The idea behind this single domain approach is based on the dynamical scaling hypothesis which,
in simple words, means that at any time $t$ in a coarsening system there is only a relevant
length $L(t)$. According to this assumption 
the average time $t(L)$ needed to close a domain of initial size $L$ in such system scales as $L^z$.
Here we adopt the stratagem is to compute $t(L)$
with a single domain. Since $R$ is the initial size of such domain we indicate 
with $t(R)$ this quantity and use $t(R)\sim R^z$ as a proxy to $t(L)$. 
This, of course, implies another assumption, namely that the shrinkage of a domain
is not significantly influenced by the presence of the others, so that a single domain configuration suffices.
Previous studies~\cite{CorLipZan08,CLP_epl,CLP_review,libro} proved that such an approach works well
because as long as $\sigma>0$ the interaction is integrable, i.e. $\sum_{\vec r} J(r)$ is finite.
This means that the interaction between spins at distance larger than $L$, in the asymptotic stage
when $L$ has grown large, is negligible.

\subsection{Numerical results}

\begin{figure}[h]
\vspace{1.5cm}
  \centering
    \rotatebox{0}{\resizebox{.45\textwidth}{!}{\includegraphics{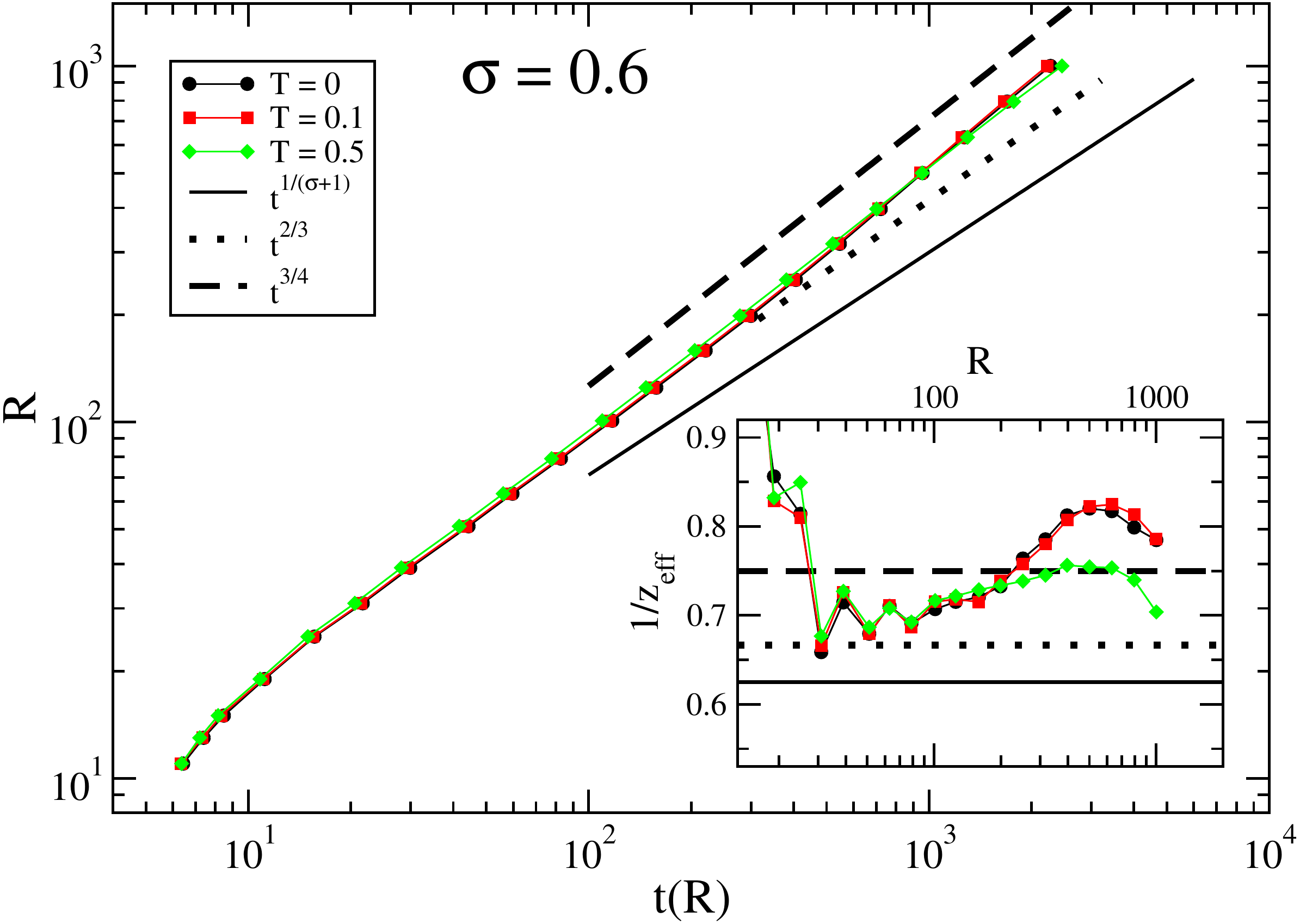}}}
  \rotatebox{0}{\resizebox{.45\textwidth}{!}{\includegraphics{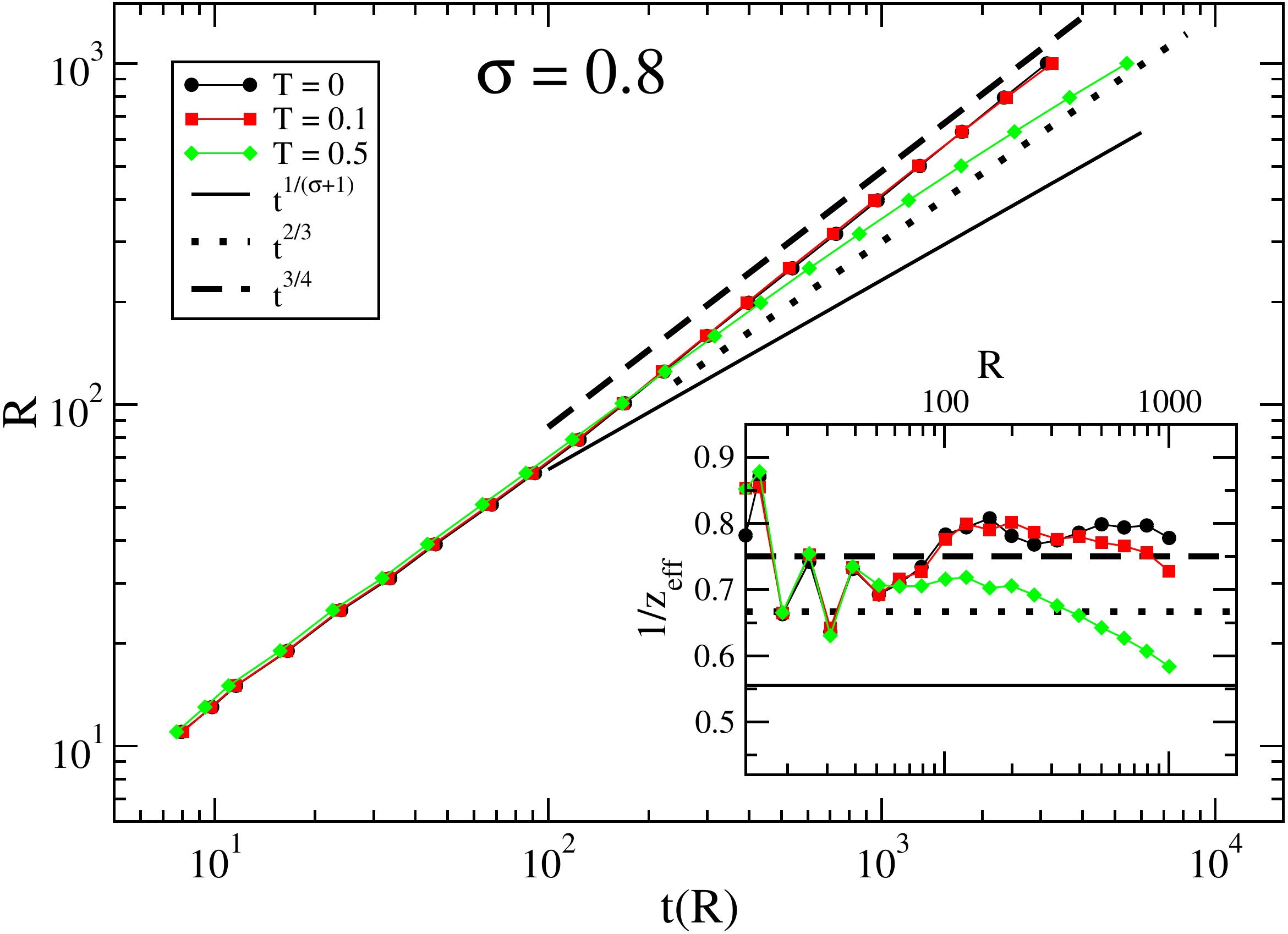}}}
    \rotatebox{0}{\resizebox{.45\textwidth}{!}{\includegraphics{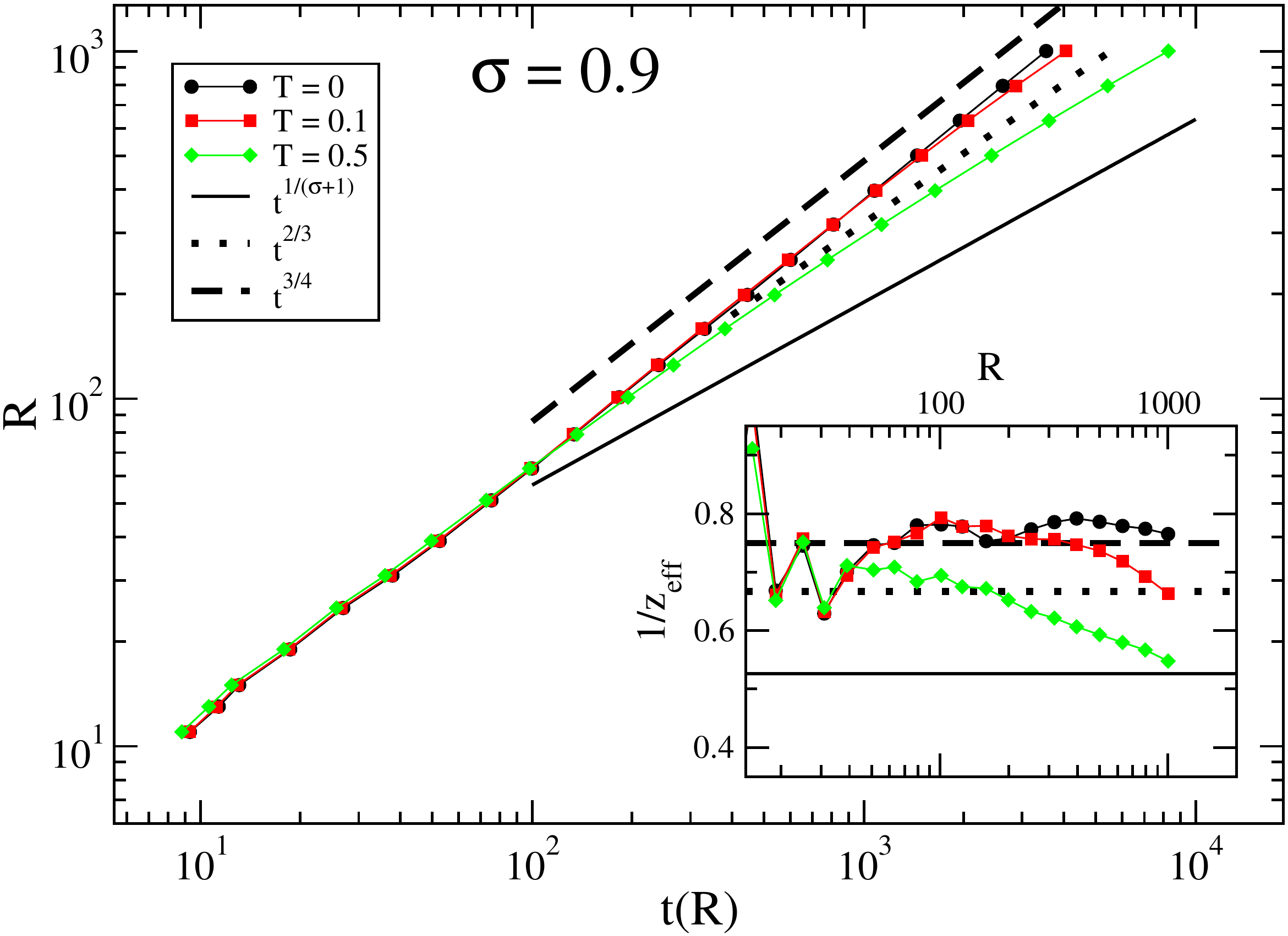}}}
      \rotatebox{0}{\resizebox{.45\textwidth}{!}{\includegraphics{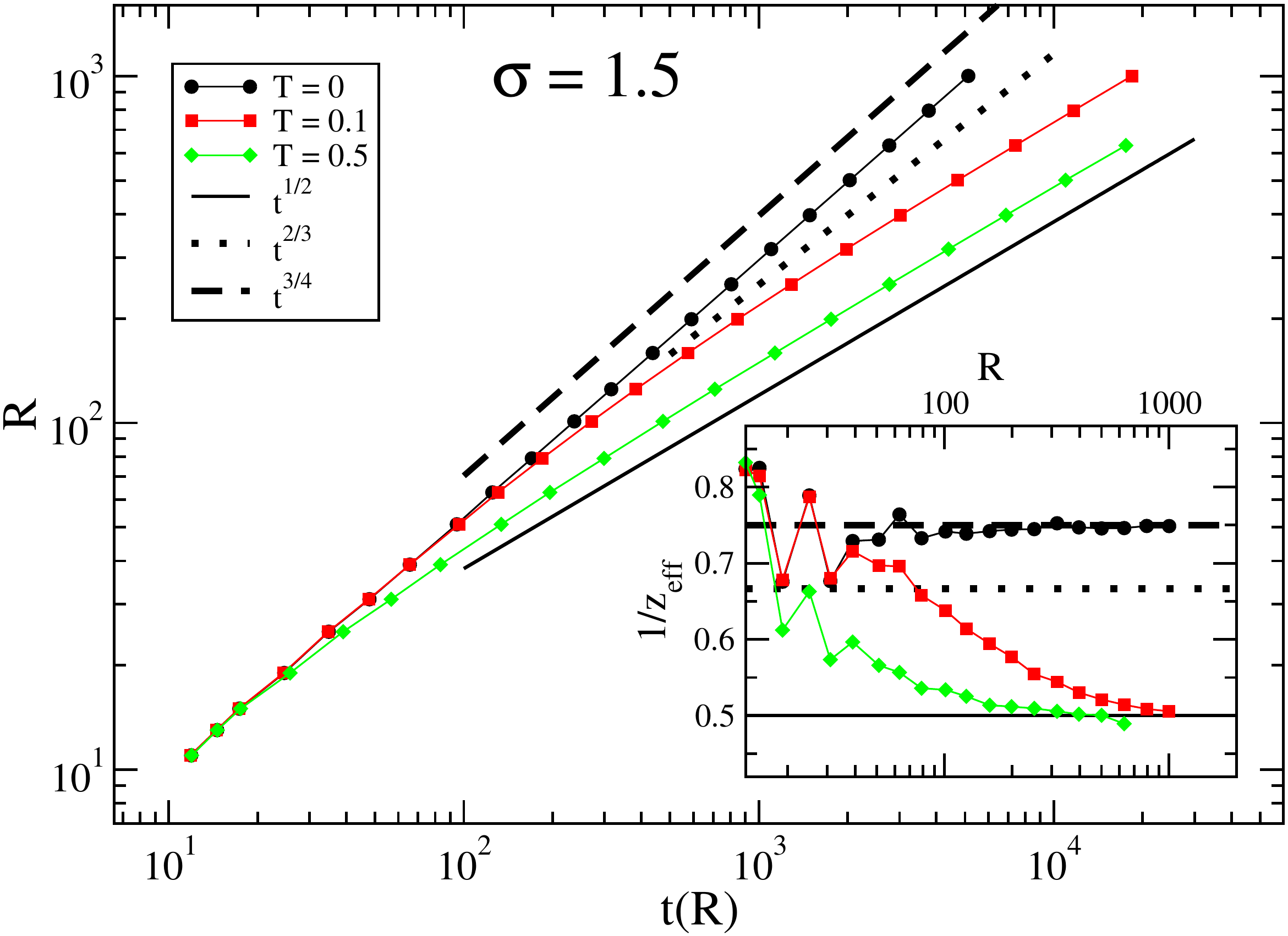}}}
          \rotatebox{0}{\resizebox{.45\textwidth}{!}{\includegraphics{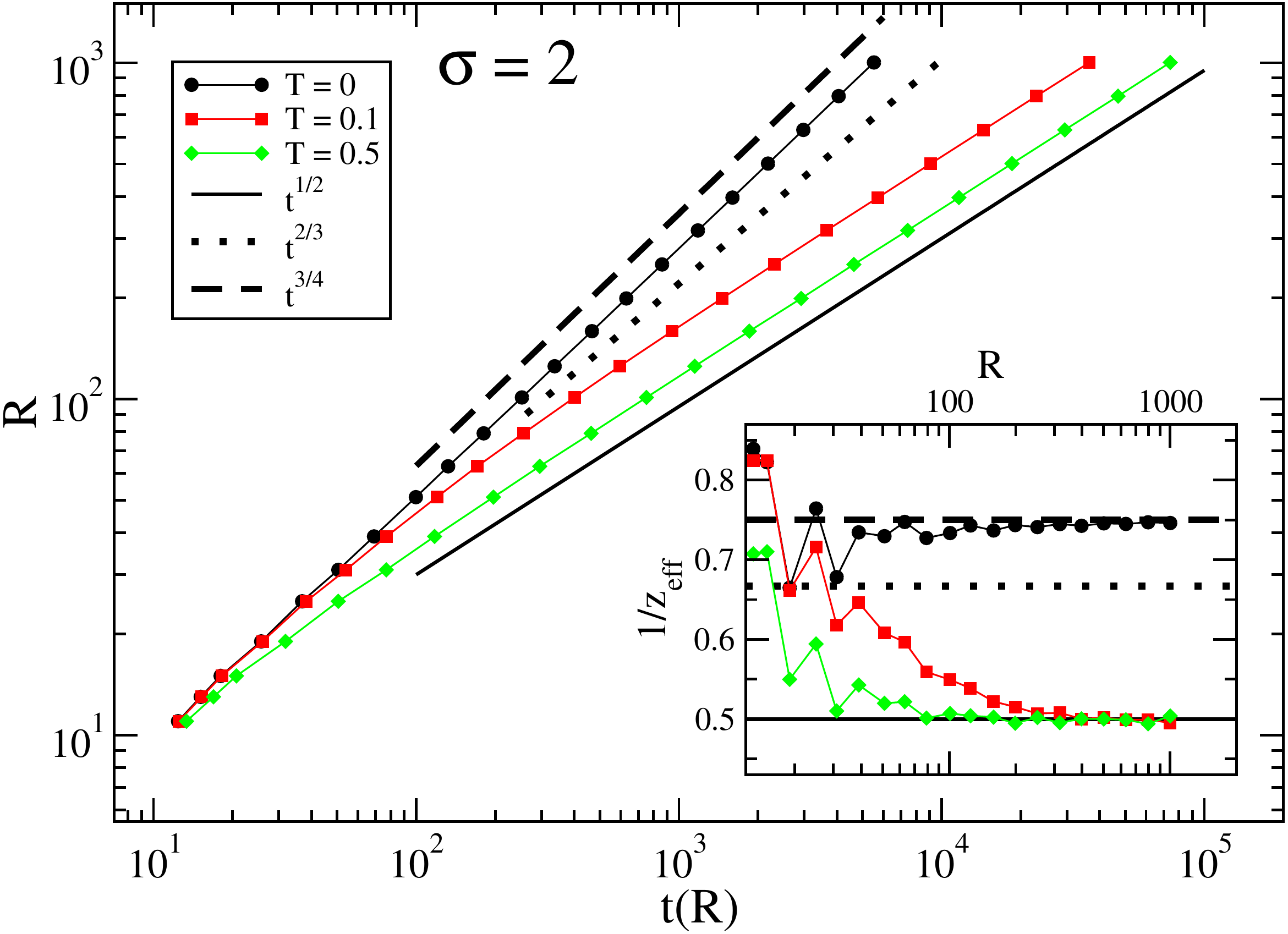}}}
    \rotatebox{0}{\resizebox{.45\textwidth}{!}{\includegraphics{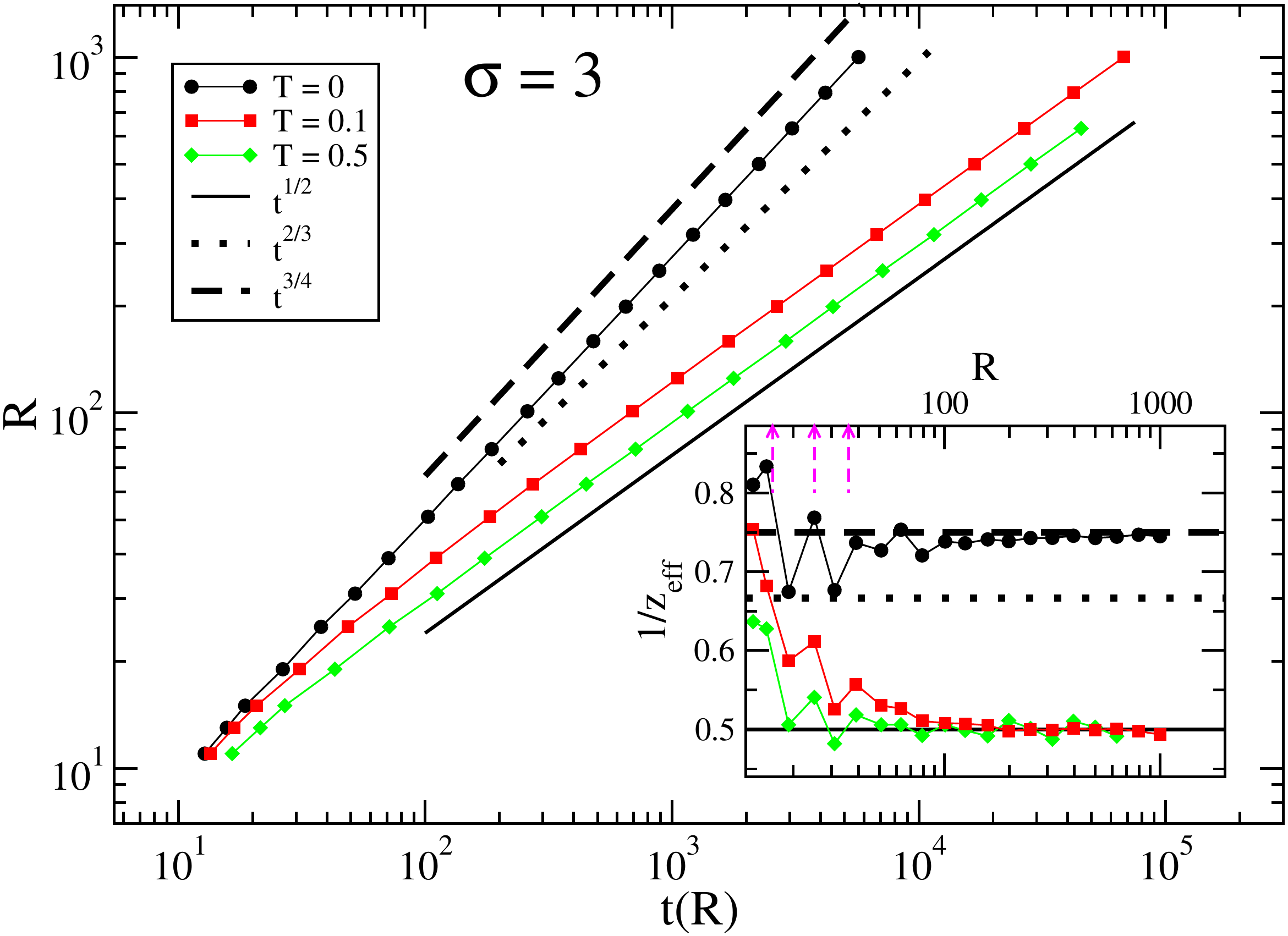}}}
  \caption{
The closure time $t(R)$ of an initially circular bubble of diameter $R$ (on the $x$ axis) is plotted against
$R$ ($y$ axis) on a double logarithmic scale for different temperatures (see key). Different panels
correspond to different values of $\sigma$, see key. The system size is $2048^2$. Data are averaged over
$10^3$ realisations for $R\le 100$ and over $10-10^2$ realisations for larger values of $R$.
The dotted and dashed lines represent the two putative
  behaviours $t^{2/3}$ and $t^{3/4}$ expected pre-asymptotically and at late times at low temperatures respectively.  
  In the inset the effective exponent $1/z_{\rm eff}$ is plotted against $R$ on a log-linear scale. The horizontal solid, dashed and dotted lines are the 
expected asymptotic and pre-asymptotic exponents. The dashed magenta upward arrows in the panel for 
$\sigma =3$ indicate the first three values
of $R$ such that $\sqrt R$ is an integer (see text), namely $R=16, 25, 36$.}
\label{bubble}
\end{figure}

To start with, we have implemented numerical simulations of the bubble shrinkage process and we have 
computed $t(R)$ by averaging over many dynamical trajectories. The results are contained in Fig.~\ref{bubble},
where in each panel we plot $t(R)$ (on the $x$ axis) vs $R$ (on the $y$ axis). 
This choice allows us a direct comparison with Fig.~\ref{preasympt} where we plotted $L(t)$ versus $t$ for an
extended multi-domain system. For the same reason we also plot 
the effective exponent, defined as previously in Eq.~(\ref{eqeffexp}) (with the replacement $L\to R$), in the inset.

Let us start with the case $\sigma =3$ (lower right panel). Working at $T=0$ one sees that a very clean value
$1/z_{\rm eff}=0.75$ sets in for large $R$, providing ${\cal Z}=4/3$. Instead, for very small sizes the effective
exponent has a zig-zag behaviour. This is perhaps due to the fact that for such small sizes some fine geometrical 
details of the initial state become relevant. We noticed indeed that the most pronounced peaks (local maxima)
of $1/z_{\rm eff}$ correspond to values of $R=n^2$, where $n$ is an integer. 
As we will further discuss below, as shown with Eq.~(\ref{topterr}), when the bubble diameter is a perfect 
square number the largest terrace $\ell _L$ of the domain (the horizontal segment 
denoted by 1 in Fig.~\ref{pict_domain})  
is naturally an integer, therefore determining a discontinuity in the dynamical process.
This effect clearly reduces and tends to disappear with increasing $R$. 
If we neglect these {\it special} maxima, for small $R$ the effective 
exponent hits twice the value $1/z_{\rm eff}=2/3$. This might
suggest that ${\cal Z}$ toggles between a pre-asymptotic value ${\cal Z}=3/2$ and an asymptotic one
${\cal Z}=4/3$. This conjecture is going to be further supported.
The asymptotic value $4/3$ is assumed for $R\gtrsim 10^2$. Comparing with 
the coarsening data of Fig.~\ref{preasympt} we can conclude that domains of such sizes are only reached
at the very end of the coarsening simulation when finite size effects start to set in. This would explain why in the 
coarsening simulations one mostly observe an exponent in between ${\cal Z}=3/2$ and ${\cal Z}=4/3$
and only a hint of the convergence to the late one ${\cal Z}=4/3$ can be observed. Looking
now still in Fig.~\ref{bubble} for $\sigma =3$, but focusing on the data at the finite temperature
$T=0.5$, one sees that the effective exponent attains the value of Bray and Rutenberg, $1/z=0.5$.
This result attests the correctness of the single domain configuration to extract the exponent $z$.

We can now pass to the other panels of Fig.~\ref{bubble}, corresponding to smaller values of $\sigma$.
At finite $T$ one concludes that there is a 
crossover phenomenon at a value $R_{\rm cross}(T,\sigma)$, where the exponent $z$
changes from ${\cal Z}$ to the Bray and Rutenberg value (for $\sigma =0.6$, $R_{\rm cross}(T,\sigma)$ is 
probably too large to observe the crossover). 
At $T=0$
the same value ${\cal Z}=4/3$ is neatly observed also for $\sigma =1.5$ and $2$. For values of $\sigma $ smaller than one
the determination of such exponent turns out to be less precise and there is the tendency to observe
larger values of $1/z_{\rm eff}$ upon decreasing $\sigma$.  This could be due to a stability effect 
that we will discuss below. The data for the effective exponent at $T=0$ are summarised in the lower panel
of Fig.~\ref{summary_exp}.

\subsection{Stability of the bubble}
\label{sec.stab}

After having presented the results of the simulations of the bubble shrinkage, which provide a determination
of the growth law in a semi-quantitative agreement with what observed in the full coarsening model, we turn now to a study
of the microscopic kinetic mechanisms producing the closure process, in order to gain some better understanding.
We start by discussing the stability properties of an interface between two regions with differently aligned spins.
The zero-$T$ stability of a (positive, let's say) spin depends on the total field $h$ acting on it. 
If $h>0$ the spin is stable (meaning that it cannot flip), if $h<0$ the spin is unstable, if $h=0$ the spin can also flip
but this usually corresponds to a neutral equilibrium, typical of the nn model (see below) but 
not of the long-range one, because
a perfect compensation of extended interactions leading to $h=0$ is almost impossible.

In Fig.~\ref{stab_interf} we consider different types of discrete interfaces and
we refer the reader to its detailed caption.
A straight interface parallel to a lattice axis, see $(a)$, is clearly stable because all spins parallel
to a given interface spin, denoted with a $\bullet$, block its reversal while all other spins, above and below such line,
perfectly compensate. This is true for long-range coupling but also for nn coupling:
in the former case all full grey circles block $\bullet$, in the latter case only its two neighbouring ones.
When the interface contains kinks, either keeping a constant slope (i.e. all the steps have the same length) as in $(b)$ or acquiring a finite
curvature as in $(c,d)$, things are different and more complicated. 
In the nn case kinks do not interact and each of them can move in both directions even at $T=0$. 
With long-range interactions the picture is completely different and we must distinguish between the two
sides of the kink. Let us always focus on the $\bullet$ spins. If the interface has a constant slope it is stable,
this is shown in $(b)$ (see discussion in the caption). Notice that this is true irrespectively of the
value of $\sigma$.
If the interface has a finite curvature as in $(c)$, $\bullet$ spin is unstable while the spin
at its right is stable, meaning that the kink can move uphill (to the left) only. This also occurs
for any $\sigma$.
In $(d)$ the interface has a constant slope locally around $\bullet$, i.e. terraces locally have the same length,
but further uphill and downhill terraces are longer and shorter respectively.
In this case there are short-distance stabilising interactions (due to parallel uncompensated spins)
and longer-distance destabilising interactions (due to anti-parallel uncompensated spins):
the global effect depends on the details of the interface and on the value of $\sigma$:
it is likely that stability prevails for large $\sigma$ and instability
prevails for small $\sigma$. Finally, in $(e)$ it is shown that also the edge spins of a top terrace are unstable.

Things are even more complicated when considering interface spins not at a kink.
As a matter of fact in the limit of small $\sigma$ even bulk spins can flip.
In particular, considering a circular domain of size $R$, the central spin is unstable up to a critical size 
$R=R_c\sim 2^{1/\sigma}$. Indeed the spin can't flip when the stabilising interaction 
$\sim \int _{r<R}d\vec r \,r^{-(2+\sigma)}$ prevails over the destabilising one 
$\sim \int _{r>R}d\vec r \,r^{-(2+\sigma)}$ produced by the anti-aligned spins outside the domain,
which gives the above mentioned result. Notice that in our bubble shrinkage simulations, in order to
have a reliable asymptotic (i.e. large $R$) determination of $t(R)$ one has to consider $R\gg R_c$, 
otherwise there is a correction lowering $t(R)$ (because the domain shrinks to zero almost immediately
as soon as $R$ crosses $R_c$). Since for $\sigma \to 0$ it is $R_c\to \infty$, this is the possible explanation
of the observation made, regarding Fig.~\ref{bubble}, that decreasing $\sigma $ one observes values of 
$1/{\cal Z}$ larger than $3/4$. In the next Section~\ref{Z_shape} we will clearly show the role of
bulk spin flips.

\begin{figure}[h]
\includegraphics[width=0.8\textwidth]{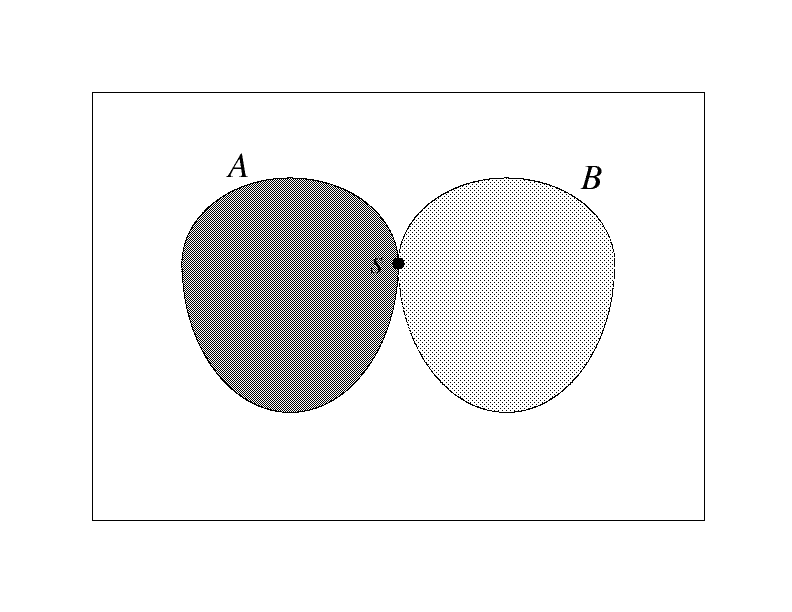}
\caption{$A$ (dark grey) is a general compact (negative) domain in a sea of positive spins.
In any point of its surface (where spin $s$ is located) we can imagine drawing the
mirror domain $B$ (light grey) tangent to it.
For symmetry reasons the field produced by the negative domain $A$ on spin $s$ is exactly compensated
by the field produced by the positive domain $B$, leaving the remaining white region, which is positively
magnetised and which therefore induces a constant drift favouring the closure of domain $A$.}
\label{continuum}
\end{figure}

We conclude this part by stressing that our results are strongly related to the discreteness of the lattice.
In a continuum medium any surface spin of a compact bubble would feel a field favouring the closure of the bubble itself
see Fig.~\ref{continuum}, because the field produced by spins within the bubble is always
compensated by the field due to the mirror domain. Uncompensated spins are anti-parallel to bubble spins
therefore inducing its closure. At zero temperature this uniform driving force would give
a trivial dynamical exponent, $z=1$.

\subsection{Simplified dynamics and determination of ${\cal Z}$}
\label{Z_shape}

In Sec.~\ref{sec.stab} we have discussed at a semi-quantitative level 
the role of bulk spin flips which produce finite-size effects 
whose importance increases with decreasing $\sigma$. The behaviour of interfacial spins with
three aligned neighbours and the fourth pointing in the opposite direction (namely spins on a locally
flat interface) is more difficult to be investigated. However we can imagine that, at least at a qualitative level,
they should behave similarly to the bulk ones, in that they are stable for large $\sigma $ and bubble sizes $R$, and
become unstable upon decreasing $R$, particularly for small $\sigma$. According to this reasoning, these spins
together with the bulk ones introduce finite bubble size effects that are more severe for small values of $\sigma$.
As already mentioned, we believe these finite size effects to be responsible for the effective exponent
$z_{\rm eff}$ somewhat larger than $4/3$ observed for small $\sigma $ in Fig.~\ref{bubble} and in the lower panel of
Fig.~\ref{summary_exp}.   
In order to avoid them, here we offer simulations where
flips in the bulk and with three aligned neighbours are forbidden.

\begin{figure}[h]
  \vspace{1.5cm}
 \centering
\rotatebox{0}{\resizebox{.85\textwidth}{!}{\includegraphics{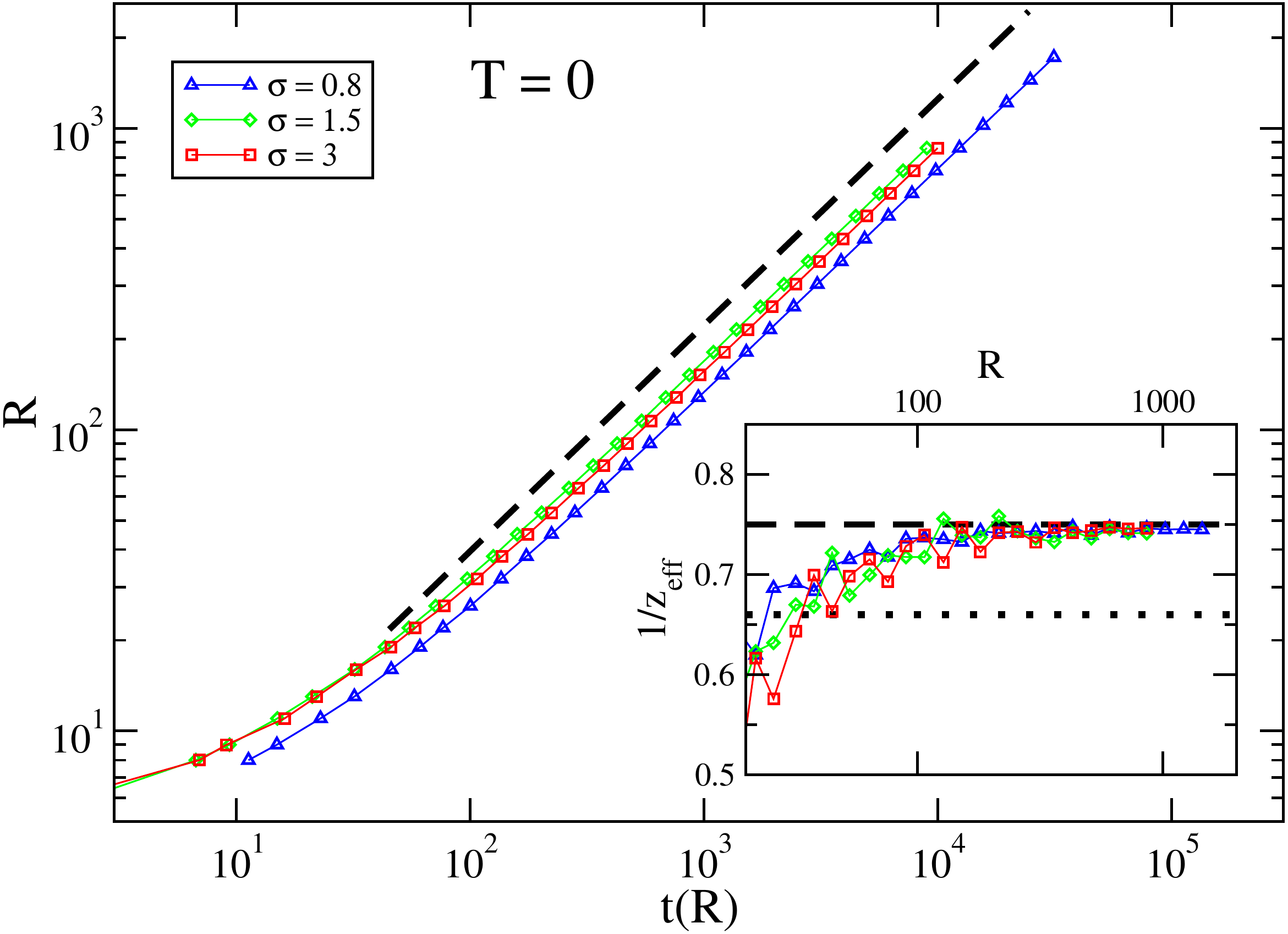}}}
\caption{The closure time $t(R)$ of a bubble of size $R$ (on the $x$ axis) is plotted against
  $R$ ($y$ axis) on a double logarithmic scale, at $T=0$ and using the simplified dynamics where only corner spins can flip. Data are averaged over $100$ realisations.
The dashed line represents the behaviour $t^{3/4}$.  
In the inset the effective exponent $1/z_{\rm eff}$ is plotted against $R$ on a log-linear scale. The horizontal dotted and dashed lines are the expected pre-asymptotic and asymptotic exponents $2/3$ and $3/4$. }
  \label{bubble_simple}
\end{figure}

Results of simulations conducted with this simplified dynamics at $T=0$ are shown in Fig.~\ref{bubble_simple}.
Our data clearly prove that the dynamical
exponent with the simplified dynamics is ${\cal Z}=4/3$ for all $\sigma$.
This suggests that this exponent is universal (i.e. independent of $\sigma$) and that $4/3$ is its truly asymptotic value. 
However, according to the previous considerations, in order to see this value with a comparable evidence for small 
$\sigma$ in full simulations one should access very large bubble sizes which is not feasible with our current numerical resources.

In order to clarify the physical mechanisms leading to ${\cal Z}=4/3$ and to the pre-asymptotic value ${\cal Z}=3/2$, it is necessary to have a more detailed study of the shape of the shrinking bubble.
If we focus on a quadrant and suppose that only edge spins can flip, the dynamics
of the droplet can be reduced to that of the terraces pictorially depicted in Fig.~\ref{pict_domain}.
We indicate with $\ell^{(j)}_r$ the length of the $j$-th terrace exactly at the time step when the top terrace 
shrinks to zero, in the configuration when the bubble radius is (at that time) $r$. 

\begin{figure}[h]
  \vspace{1.5cm}
 \centering
\rotatebox{0}{\resizebox{.85\textwidth}{!}{\includegraphics{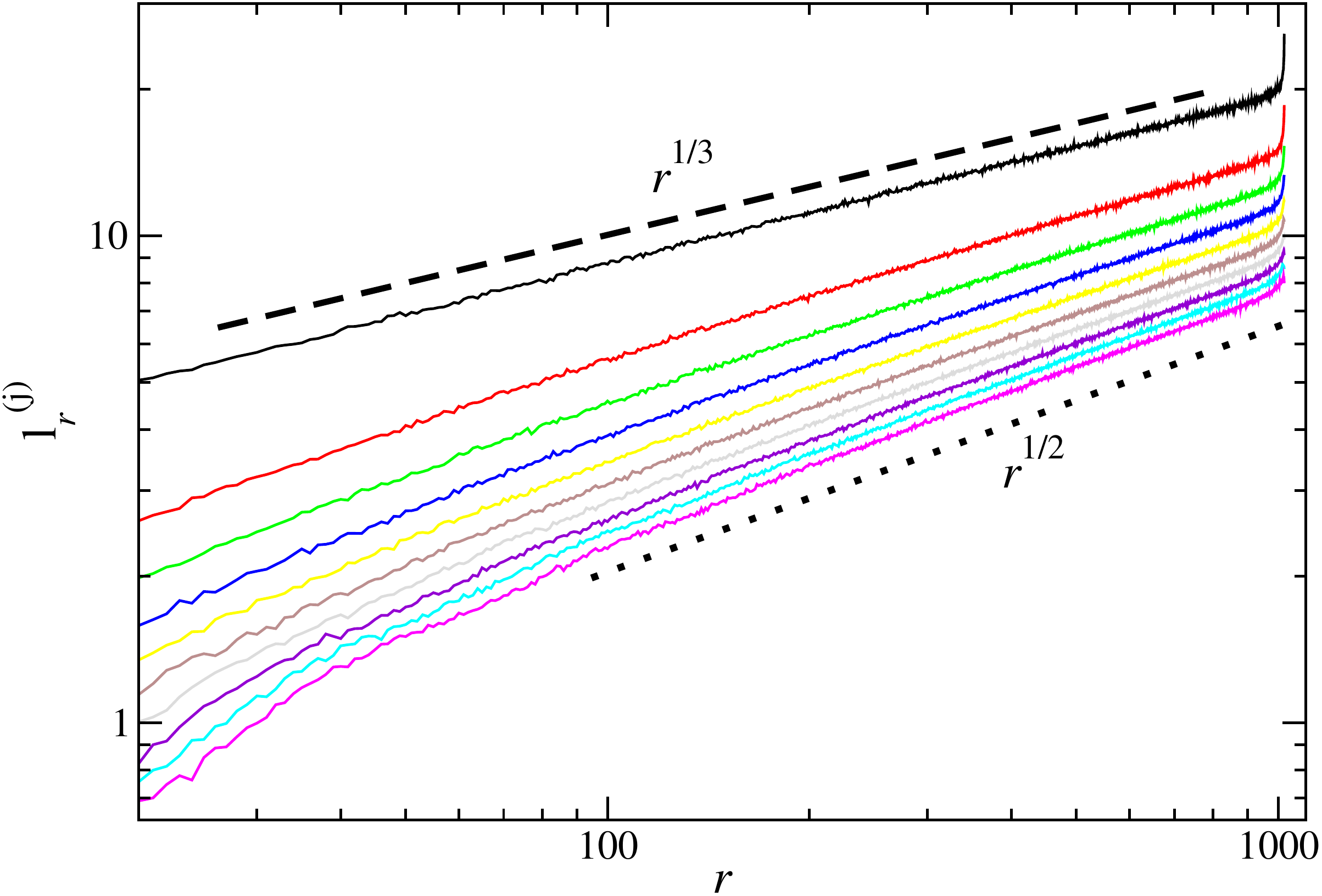}}}
\caption{The length of the $j$-th terrace $\ell_r^{(j)}$ at the closure time of the top terrace is plotted
against $r$, for $j=1-10$, from top to bottom. The dashed line is the power law $r^{1/3}$ , 
whereas the dotted line is the one $r^{1/2}$.}
  \label{terrace}
\end{figure}

\begin{figure}[h]
  \vspace{1.5cm}
 \centering
\rotatebox{0}{\resizebox{.85\textwidth}{!}{\includegraphics{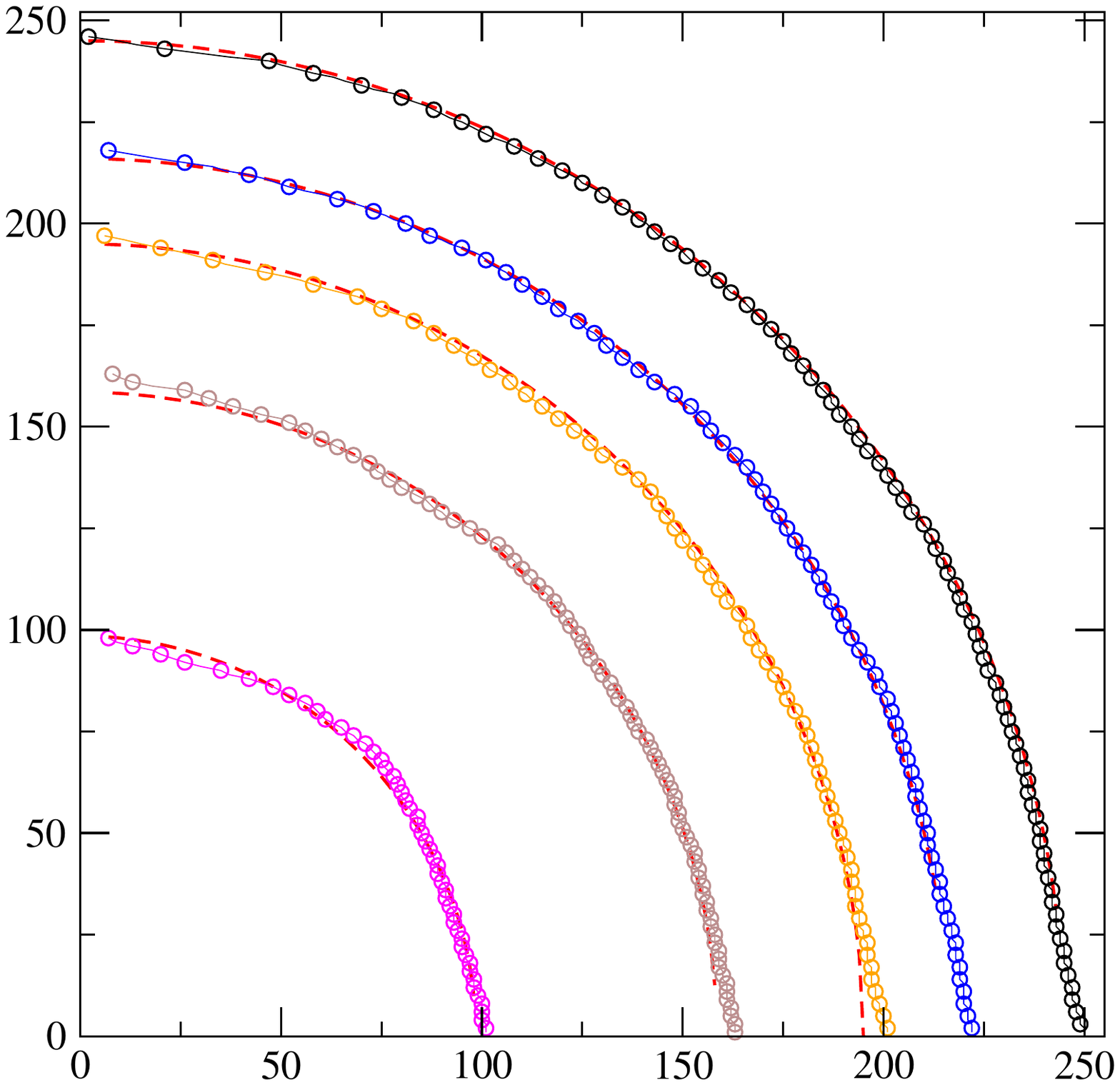}}}
\caption{The shape of the shrinking bubble with the approximated dynamics for $\sigma=0.8$. Different symbols corresponds to the droplet configuration at different times $t=10*2^i$, with $i=1,4,5,6$ increasing from top to bottom. The dashed red curve is a circle in the continuum.  }
  \label{profile}
\end{figure}

Taking into account that the kink on the upper terrace always moves ballistically, we find that the time necessary for the droplet to disappear is
\be
t(R) = \sum _{r=2}^{R} \ell_r^{(2)}+\ell_R,
\label{tl}
\ee
where $\ell_R$ is the length of the top terrace in the initial configuration.
The evaluation of $\ell_r^{(2)}$ is complicated because of the ``long-range'' interaction between terraces. An analytical expression for $\ell_r^{(2)}$ can be obtained from the observation (see Fig.~\ref{bubble_geom}) that,
independently of the initial configuration, the droplet asymptotically assumes a quasi circular shape. Assuming that the droplet is {\it exactly} circular at all times,   
the quantity $\ell_r^{(2)}$ is obtained from the equation 
of the boundary in the continuum, $x^2+y^2=r^2$, 
as the value of $x$ corresponding to $y=r-1$, 
\be
\ell_r^{(2)}=\sqrt{2r-1} \simeq \sqrt{2r}.
\label{topterr}
\ee
Inserting this expression in Eq.~(\ref{tl}) we get $t(R)\sim R^{3/2}$, hence ${\cal Z}$=3/2.
This value, as we know, is observed for small $R$, whereas for large $R$ it crosses over
to ${\cal Z}=4/3$ indicating a deviation from the circular profile of the upper terraces. 
This deviation must be attributed to the interaction between different terraces, in particular because of the drive exerted by the ballistic upper terrace transporting to some extent the lower ones. In order to be more quantitative about this,
in Fig.~\ref{terrace} we plot results for $\ell_r^{(j)}$ as a function of $r$ for $j\le 10$. 
Let us look at the region $10^2<r<10^3$. In this sector one sees that the size of the 
innermost of the considered terraces, namely $j=10$ (lowest curve, magenta) grows as $r^{1/2}$. Recalling Eq.~(\ref{topterr}), and observing that also $\ell ^{(j)}_r\propto \sqrt {r}$, for $r \gg j$, 
this is what one would expect for a circular domain indicating that, sufficiently
far from the top of the domain, its shape is circular. However, upon decreasing $j$, that is considering 
higher terraces, this slope gradually decreases until, for $j=2$, the top terrace behaves as 
$\ell_r^{(2)}\sim r^{1/3}$. Using this result in Eq.~(\ref{tl}) 
one gets from the behaviour of this  terrace the correct asymptotic value ${\cal Z}=4/3$.
This shows that the origin of this number must be traced back to the non trivial geometry of the domains induced by 
non local interactions between terraces. The actual shape of the domain during its shrinkage is shown in Fig.~\ref{profile},
where one can see deviations from circularity in the top and rightmost part of the bubble. Notice that such 
deviations are hardly visible to the naked eye, however their effect on ${\cal Z}$ is important enough to change
its value from $3/2$ to $4/3$. 

To complete the discussion of Fig.~\ref{terrace}, let us notice that there is a steep increase at $r\gtrsim 10^3$
and also that the algebraic behaviour of the curves is spoiled for small values of $r$. The former effect is due to the fact that the simulation is initialised with a domain of size $\gtrsim 1000$ and it takes some time for the dynamical process to build the domain's shape and to enter the scaling stage. The latter effect occurs because, in the very late stage of the process when the domain is very small, the scaling properties are lost because there are not enough interacting terraces to produce a many-body phenomenon.

\section{Conclusions}
\label{concl}

In this paper, we have shown that during the zero-temperature relaxation dynamics of the two-dimensional long-range Ising model a new dynamical regime appears.  This is characterised by an exponent ${\cal Z}$ that takes a pre-asymptotic value 
${\cal Z}=3/2$ and an asymptotic one ${\cal Z}=4/3$. These values are strongly related to the interplay between the discrete nature of the lattice and the long-range nature of the interaction. In fact, if interactions are short-range (nn model)
we would have $z_{\rm cd}=2$ both in a discrete lattice and in a continuum picture (see Sec.~\ref{intro}).
On the other hand, a continuum picture with long-range interactions would trivially give $z=1$ (see Fig.~\ref{continuum} and related discussion at the end of Sec.~\ref{sec.stab}).

The pre-asymptotic value ${\cal Z}=3/2$ [see Eqs.~(\ref{tl}-\ref{topterr})] is simply understandable. For this, we use a discrete approximation of a circular bubble and assume that its shrinkage dynamics is limited by the closure of the largest terraces. However, Fig.~\ref{terrace} shows that the largest terraces scale as $r^{1/3}$ rather than as $r^{1/2}$, as expected for a circular bubble. This results in the true asymptotic value ${\cal Z}=4/3$. The reason for such a behaviour, see Sec.~\ref{Z_shape}, can be associated with the long-range character of the interaction and the necessity to locally preserve a convex shape of the bubble in order for the dynamics to proceed. Apart from this phenomenological understanding of the mechanism producing the asymptotic value ${\cal Z}=4/3$, a rigorous analytical derivation has not yet been possible.

Given the paucity of analytic approaches, further analysis is confined to better numerical simulations of the multi-domain system. To be useful, these must go beyond the time-scales and the statistical accuracy of those presented in this paper.  As we estimate below, this seems to be possible, but entails a huge computational effort.
Fig.~\ref{summary_exp} shows that the effective exponent would reasonably approach the value $4/3$ for sizes
of order $L(t) \simeq 200-300$. Thus, one should check for its further behaviour up to, say, $L(t) \simeq 500-600$.
Fig.~\ref{preasympt} informs us that, in order to reach this point, we should go to times of order $t\simeq 10^4$. Also, from the same figure one understands that finite-size effects start to be important around $L(t) \simeq 100 \simeq {\cal L}/20$. Hence, in order to reach $L(t) \simeq 500$ without severe finite-size effects, the system should have at least a size ${\cal L} \simeq 10^4$. Putting the above facts together, we estimate a required computational effort at least 100 times larger than the one of this study, which is already quite massive. This estimate applies for the larger values of $\sigma$, for smaller values the situation is much worse!

The study of this paper is restricted to a two-dimensional square lattice. However, since we have emphasised that
lattice effects are relevant in determining the $T=0$ growth exponent, an interesting question to be still addressed is the universality of the exponent ${\cal Z}$ with respect to the lattice geometry. In this direction, numerical simulations of $d=2$ systems on different lattices (e.g., triangular or hexagonal lattices) would be useful. Similarly, the dependence on dimensionality should also be considered, although the numerical effort required would be further increased. Finally, the dynamics with extremely long-range interactions (i.e., with $\sigma \le 0$) remains a completely unexplored subject in any dimension.

Few days after submitting this article for publication, a preprint~\cite{Christiansen_arxiv} has been uploaded 
where the authors also find the exponent $3/4$ independently of $\sigma$.

\appendix*

\section{Ewald Summation Technique}
\label{Ewald}

Combining the effective interaction between two spins described by Eq.~(\ref{jsum}) in Eq.~(\ref{ham}), the Hamiltonian for the long-range Ising model takes the form:
\be
{\cal H}(\{s_i\})=-\frac{1}{2} \sum_{\vec n}{}^{'} \sum_i \sum_j \frac{1}{\vert \vec n + \vec r_{ij} \vert ^{2+\sigma}} s_is_j,
\label{hameff}
\ee
where $\vec r_{ij} = \vec r_{i} - \vec r_{j}$, and the summation over $\vec n$ accounts for the contribution of infinite imaginary copies across periodic boundaries. The prime in this summation indicates that, when $\vec n = (0,0)$, the terms with $i=j$ are excluded. In this Appendix, we will split the summation over $\vec n$ into a combination of two rapidly convergent summations. For this, we take the help of complete and incomplete gamma functions defined respectively as
\be
\Gamma (x)=\int_{0}^{\infty} t^{x-1} e^{-t} dt,
\label{Gamma}
\ee
\be
\Gamma (x,y)=\int_{y}^{\infty} t^{x-1} e^{-t} dt.
\label{incGamma}
\ee

The trick of Ewald, which was originally proposed for the Coulomb potential \cite{fs2002,ew1921}, is introduced as follows. Firstly, we use the integral of Eq.~(\ref{Gamma}) and divide the interval of integration into two parts:
\bea
J(s_i,s_j) &=& \sum_{\vec n} \frac{1}{\vert \vec n + \vec r_{ij} \vert^{2+\sigma}}=\sum_{\vec n} \frac{1}{\Gamma \left(\sigma/2+1\right)} \int_{0}^{\infty} \frac{1}{\vert \vec n + \vec r_{ij} \vert ^{2+\sigma}} t^{\frac{\sigma}{2}} e^{-t} dt \nonumber \\
&=& \sum_{\vec n} \frac{1}{\Gamma \left(\sigma/2+1\right)} \left(\int_{0}^{\alpha^2 \vert \vec n + \vec r_{ij} \vert^2} \frac{1}{\vert \vec n + \vec r_{ij} \vert ^{2+\sigma}} t^{\frac{\sigma}{2}} e^{-t} dt
+ \int_{\alpha^2 \vert \vec n + \vec r_{ij} \vert^2}^{\infty} \frac{1}{\vert \vec n + \vec r_{ij} \vert ^{2+\sigma}} t^{\frac{\sigma}{2}} e^{-t} dt \right), \nonumber \\
&=& \ I_1 + I_2,
\label{split}
\eea
where $\alpha$ is a positive real number. The separated terms, denoted by $I_1$ and $I_2$, represent the contributions of the integral over intervals $0 < t < \alpha^2 \vert \vec n + \vec r_{ij} \vert^2 $ and $\alpha^2 \vert \vec n + \vec r_{ij} \vert^2 < t < \infty$, respectively.

Looking at the second term given in Eq.~(\ref{split}), one can write with the help of Eq.~(\ref{incGamma})
\be
I_2=\frac{1}{\Gamma \left(\sigma/2+1\right)} \sum_{\vec n} \frac{1}{\vert \vec n + \vec r_{ij} \vert ^{2+\sigma}} \Gamma \left(\frac{\sigma}{2}+1,\alpha^2 \vert \vec n + \vec r_{ij} \vert ^2 \right).
\label{IInd}
\ee
This term rapidly converges as $\vert \vec n \vert$ increases, representing the short-range contribution of exchange-interaction between spins $s_i$ and $s_j$.

To simplify the first term, we use the change of variables $\rho^2 \vert \vec n + \vec r_{ij} \vert ^2 = t$,
\be
I_1=\frac{2}{\Gamma \left(\sigma/2+1\right)} \sum_{\vec n} \int_{0}^{\alpha} \rho^{\sigma+1} e^{-|\vec n + \vec r_{ij}|^2 \rho^2} d\rho.
\label{Ist}
\ee
The above term can also be made rapidly convergent by transforming to reciprocal space. We use the Poisson summation formula to obtain
\be
\sum_{\vec n} e^{-|\vec n + \vec r_{ij}|^2 \rho^2} = \frac{\pi}{{\cal L}^2 \rho^2} \sum_{\vec k} e^{-\pi^2|\vec k|^2/ \rho^2} e^{i 2\pi \vec k \cdot \vec r_{ij}},
\label{Poiss}
\ee
where the reciprocal vector $\vec k = (k_x,k_y)$ with $k_x, k_y = 0, \pm 1/{\cal L}, \pm 2/{\cal L}, ...$. As mentioned in the main text, ${\cal L}$ denotes the system size. Using the above expression in Eq.~(\ref{Ist}), and recalling Eq.~(\ref{incGamma}), the first term (\ref{Ist}) can be written as
\be
I_1=\frac{2 \pi}{{\cal L}^2 \Gamma \left(\sigma/2+1\right)} \sum_{\vec k} e^{i 2\pi \vec{k} \cdot \vec r_{ij}} \frac{1}{2}(\pi \vert \vec k \vert)^{\sigma} \Gamma \left(-\frac{\sigma}{2},\frac{\pi^2 \vert \vec k \vert ^2}{\alpha^2}\right).
\label{reciproc}
\ee
This term rapidly becomes negligible as $\vert \vec k \vert$ increases, representing the long-range contribution of exchange interactions. Combining Eq.~(\ref{IInd}) and Eq.~(\ref{reciproc}) in Eq.~({\ref{split}}) provides the essence of the Ewald summation technique. One should also note that the summations over $\vec n$ and $\vec k$ in Eq.~(\ref{IInd}) and Eq.~(\ref{reciproc}) respectively are conditionally convergent, i.e., the convergence of the summations depend on the order of adding terms in the summations. The best way is to sum spherically over $\vert \vec n \vert$ and $\vert \vec k \vert$.

The auxiliary parameter $\alpha$ determines the speed of convergence of summations over $\vert \vec n \vert$ and $\vert \vec k \vert$. We have taken the value of $\alpha = 2/{\cal L}$, also chosen in recent paper \cite{PhysRevE.95.012143}, which allows us to truncate the summation to $\vert n\vert \le 5 {\cal L}$ in $I_2$ (Eq.(\ref{IInd})) and to $\vert k \vert \ge 5/{\cal L}$ in $I_1$ (Eq.(\ref{reciproc})). To calculate complete and incomplete gamma functions in the numerical simulations, we have used the Fortran interface of GNU scientific library (FGSL) in gfortran.

\bibliography{lrim2d}

\acknowledgments{E. Lippiello and P. Politi acknowledge
  support from the MIUR PRIN 2017 project 201798CZLJ}

\end{document}